\documentclass[preprint,preprintnumbers,amsmath,amssymb,floatfix,nofootinbib]{revtex4}
\usepackage{epsf}
\usepackage{graphicx}

\def\lsim{\mathrel{\rlap{\lower4pt\hbox{\hskip1pt$\sim$}}
    \raise1pt\hbox{$<$}}}
\def\gsim{\mathrel{\rlap{\lower4pt\hbox{\hskip1pt$\sim$}}
    \raise1pt\hbox{$>$}}} 

\newcommand{\slashed}[1]{\setbox0=\hbox{$#1$}   
   \dimen0=\wd0                                     
   \setbox1=\hbox{/} \dimen1=\wd1                   
   \ifdim\dimen0>\dimen1                            
      \rlap{\hbox to \dimen0{\hfil/\hfil}}          
      #1                                            
   \else                                            
      \rlap{\hbox to \dimen1{\hfil$#1$\hfil}}       
      /                                             
   \fi}                                             %

\begin{document}          

\vskip 2in
\hskip 5.2in \hbox{\bf YITP-SB-11-11}

\title{\vspace*{0.7in}
 SUSY QCD corrections to Higgs-b Production:  Is the $\Delta_b$ Approximation
Accurate?}

\author{S.~Dawson$^{a}$, C.~B.~Jackson$^{b}$, P.~Jaiswal$^{a,c}$}

\affiliation{
$^a$Department of Physics, Brookhaven National Laboratory, 
Upton, NY 11973, USA \\
$^{b}$ Physics Department, University of Texas, Arlington, Texas \\
$^{c}$Yang Institute for Theoretical Physics, Stony Brook University, Stony Brook, NY 11790, USA
\vspace*{.5in}}

\begin{abstract}
The associated production of a Higgs boson with a $b$ quark is a discovery channel for the
lightest MSSM neutral Higgs boson.  We consider the SUSY QCD contributions 
from squarks and gluinos and discuss the decoupling properties of
these effects. A detailed comparison of our exact ${\cal O}(\alpha_s)$
results with those of a widely used effective Lagrangian approach, the
$\Delta_b$ approximation, is presented.  The $\Delta_b$ approximation
is shown to accurately reproduce the exact one-loop SQCD
 result to within a few percent over
a wide range of parameter space.

\end{abstract}

\maketitle
\newpage

\section{Introduction}

Once a light Higgs-like particle is discovered it will be critical to determine if it is the Higgs Boson predicted by the Standard Model.  The
minimal supersymmetric Standard Model  (MSSM) presents a
comparison framework in which
 to examine the properties of a putative Higgs candidate.  The MSSM Higgs 
sector contains $5$ Higgs bosons--$2$ neutral
bosons, $h$ and $H$, a pseudoscalar boson, $A$, and $2$ charged 
bosons, $H^\pm$. At the tree level the theory
 is described by just $2$ parameters, 
which are conveniently chosen to be $M_A$, the mass of the pseudoscalar boson,
 and $\tan\beta$, the ratio of
vacuum expectation values of the $2$ neutral Higgs bosons. 
Even when radiative corrections are included, the theory is highly 
predictive\cite{Djouadi:2005gj,Gunion:1989we,Carena:2002es}.
 
In the MSSM, the
production mechanisms for the Higgs bosons can be significantly different from
in the Standard Model. For large values of $\tan\beta$, the heavier Higgs
bosons, $A$ and $H$,
are predominantly produced in association with $b$ quarks.  
Even for 
$\tan\beta\sim 5$, the production rate in association with $b$ quarks is
similar to that from gluon fusion for $A$ and $H$
production\cite{Dittmaier:2011ti}. For the lighter Higgs
boson, $h$, for
$\tan\beta \gsim 7$ the dominant production mechanism 
 at both the Tevatron and the LHC
 is production with $b$ quarks for light $M_A$ ($\lsim 200~GeV$), where 
the $b {\overline b} h$ coupling is enhanced .
Both the Tevatron\cite{Benjamin:2010xb} 
and the LHC experiments\cite{Chatrchyan:2011nx}
have presented limits Higgs production in association
with $b$ quarks, searching for the 
decays $h\rightarrow \tau^+\tau^-$ and $b \overline{b}$\footnote{The
expected sensitivities of ATLAS and CMS to $b$ Higgs associated
production are described in Refs. \cite{Aad:2009wy,Ball:2007zza}.}.
  These limits 
are obtained in the context of the MSSM are sensitive 
to the $b$-squark and gluino
loop corrections which we consider here.

The rates for $bh$ associated production at
the LHC and the Tevatron 
have been extensively studied\cite{Dawson:2005vi,Campbell:2004pu,Maltoni:2003pn,Dawson:2004sh,Dittmaier:2003ej,Dicus:1998hs,Dawson:2003kb,Maltoni:2005wd,Campbell:2002zm,Carena:2007aq,Carena:1998gk} and the NLO QCD correction
are well understood, 
both in the $4$- and $5$- flavor number parton 
schemes\cite{Dawson:2005vi,Campbell:2004pu,Maltoni:2003pn}.  
In the $4$- flavor number scheme, the lowest order processes 
for producing a Higgs boson and a $b$ quark are $gg\rightarrow
 b {\overline b}h$ and 
$ q {\overline q}\rightarrow b {\overline b} 
h$\cite{Dittmaier:2003ej,Dawson:2004sh,Dicus:1998hs}.  In the $5-$ flavor
number scheme, the lowest order process is $b g \rightarrow b h$
 (${\overline {b}} g \rightarrow {\overline {b}}h$).
The two schemes represent different orderings of perturbation theory and
  calculations
in the two schemes produce rates which are in qualitative 
agreement\cite{Campbell:2004pu,Dittmaier:2011ti}.  
In this paper, we use the $5$-flavor number scheme for simplicity. 
The resummation of threshold logarithms\cite{Field:2007ye}, 
electroweak corrections\cite{Dawson:2010yz,Beccaria:2010fg}
 and SUSY QCD corrections\cite{Dawson:2007ur} have also been
computed for $bh$ production in the $5-$ flavor number scheme.
  
Here, we focus on the role of squark and gluino loops.  
The properties of the SUSY QCD corrections to the $b {\overline b}
 h$ vertex, both
for the decay $h\rightarrow b {\overline b}
$\cite{Dabelstein:1995js,Hall:1993gn,Carena:1999py,Guasch:2003cv}
and the production, 
$b {\overline b}\rightarrow h$\cite{Haber:2000kq,Harlander:2003ai,Guasch:2003cv,Dittmaier:2003ej}, were computed
long ago.  
The contributions from $b$ squarks and gluinos to the
lightest MSSM Higgs boson
mass are known at $2$-loops\cite{Heinemeyer:2004xw,Brignole:2002bz}, 
while the $2$-loop
SQCD contributions to the $b{\overline b}h$ vertex
 is known in the limit in which the
Higgs mass is much smaller than the squark and gluino 
masses\cite{Noth:2010jy,Noth:2008tw}.
The contributions of squarks and gluinos to the 
on-shell $b {\overline b}
h$ vertex are non-decoupling for heavy squark and gluino masses and
decoupling is only achieved when the pseudoscalar
mass, $M_A$, also becomes large.

An effective Lagrangian approach, the $\Delta_b$ 
approximation\cite{Carena:1999py,Hall:1993gn},
can be used to approximate 
the SQCD contributions to 
the on-shell $b {\overline b}h$ vertex
and to resum the $(\alpha_s  \tan\beta/M_{SUSY})^n$ enhanced terms.
The numerical accuracy of the $\Delta_b$ effective Lagrangian approach
has been examined for a number of cases.  
The $2-$loop contributions to
the lightest MSSM Higgs boson mass 
of ${\cal O}(\alpha_b\alpha_s)$ were computed
in Refs. \cite{Heinemeyer:2004xw} and \cite{Brignole:2002bz}, 
and it was found that the majority of these corrections
could be absorbed into a $1-$loop 
contribution by defining an effective  $b$ quark
 mass using the $\Delta_b$ approach. 
The sub-leading contributions to the Higgs
boson mass (those not absorbed into $\Delta_b$)
are then of ${\cal O}(1~GeV)$.  
The $\Delta_b$ approach  also yields an excellent
approximation to the SQCD corrections for
 the decay process $h\rightarrow b
{\overline b}$\cite{Guasch:2003cv}.  
It is particularly interesting
to study the 
accuracy of the
$\Delta_b$ approximation 
for production processes where one of the $b$ quarks is off-shell.
The SQCD contributions from squarks and gluinos to the 
inclusive Higgs production rate in association with $b$ quarks has been
studied extensively in the 4FNS in 
Ref. \cite{Dittmaier:2006cz}, where the the lowest order
contribution is $gg\rightarrow b {\overline b}h$.  
In the 4FNS, the inclusive
cross section including the exact 1-loop SQCD corrections is reproduced
to within a few percent using the $\Delta_b$ approximation.  However,
the accuracy of the $\Delta_b$ approximation for the MSSM
neutral Higgs boson 
production in the
5FNS has been studied for only a small set of MSSM parameters in 
Ref. \cite{Dawson:2007ur}.
The major new result of this paper is a detailed study of
the accuracy of the $\Delta_b$ approach in the 5FNS 
for the $bg\rightarrow bh$ production process.  In this case,
one of the $b$ quarks is off-shell and there are contributions which are not
contained in the effective Lagrangian approach. 

The plan of the paper is as follows:  Section $2$ contains a brief review
of the MSSM Higgs and $b$ squark  sectors 
and also a review of the effective Lagrangian approximation.  
The calculation of
Ref. \cite{Dawson:2007ur} is summarized in Section 2. 
We include
  SQCD contributions
to  $bh$ production which are enhanced by $m_b \tan\beta$ which were omitted
in Ref. \cite{Dawson:2007ur}. 
Analytic results for the SQCD corrections
to $bg\rightarrow bh$ in the extreme mixing scenarios in the $b$ squark
sector  are presented 
 in Section 3.  
Section 4 contains
numerical results for the $\sqrt{s}=7$ TeV
LHC.  Finally, our conclusions are summarized in Section 5.
Detailed analytic results are relegated
to a series
of appendices.

\section{Basics}

\subsection{MSSM Framework}

In the simplest version of the MSSM there are two Higgs doublets,
$H_u$ and $H_d$,
 which break the electroweak
symmetry and give masses to the $W$ and $Z$ gauge bosons.  The neutral Higgs boson masses are given
at tree level by,
\begin{equation}
M_{h,H}^2={1\over 2} \biggl[M_A^2+M_Z^2\mp\sqrt{(M_A^2+M_Z^2)^2-4M_A^2M_Z^2\cos^2 2\beta}\biggr]
\, ,
\label{mhtree}
\end{equation}
and the angle, $\alpha$, which diagonalizes the neutral Higgs mass is
\begin{equation}
\tan 2\alpha=\tan 2\beta
\biggl( {M_A^2+M_Z^2\over M_A^2-M_Z^2}\biggr) \, .
\label{alphatree}
\end{equation}
In practice, the relations of Eqs.~\ref{mhtree} and \ref{alphatree}
receive large radiative corrections which must be taken into account
in numerical studies.  We use the program 
FeynHiggs\cite{Heinemeyer:1998yj,Degrassi:2002fi,Heinemeyer:1998np} to generate the
Higgs masses and an effective mixing angle, $\alpha_{eff}$, which 
incorporates higher order effects.

The scalar partners of the left- and right- handed $b$ quarks, ${\tilde b}_L$
and ${\tilde b}_R$, are not mass eigenstates, but mix according to,
\begin{equation}
L_M=-({\tilde b}^*_L, {\tilde b}^*_R)M_{\tilde b}^2 \left(
\begin{array}{c}
{\tilde b}_L \\
{\tilde b}_R
\end{array}
\right)\, .
\end{equation}
 The ${\tilde b}$ squark  mass matrix is,
\begin{equation}
M_{{\tilde b}}^2=\left(
\begin{array}{cc}
{\tilde m}_L^2 & m_b X_b\\
m_b X_b & {\tilde m}_R^2\\
\end{array}
\right)\, ,
\label{squark}
\end {equation}
and we define,
\begin{eqnarray}
X_b&=& A_b-\mu\tan\beta\nonumber \\
{\tilde m}^2_L&=&
{ M}_Q^2 +m_b^2+M_Z^2\cos 2\beta (I_3^b-Q_b\sin^2\theta_W)\nonumber \\
{\tilde m}^2_R&=&
{M}_D^2 +m_b^2+M_Z^2\cos 2\beta Q_b\sin^2\theta_W\, .
\end{eqnarray}
${ M}_{Q,D}$ are the soft SUSY breaking masses,
$I_3^b=-1/2$, and $Q_b=-1/3$.  The parameter $A_b$ is the trilinear scalar coupling of the soft supersymmetry breaking Lagrangian and $\mu$ is the Higgsino mass parameter.
The $b$ squark mass eigenstates are
${\tilde b}_1$ and ${\tilde b}_2$ and define
 the $b$-squark mixing angle, ${\tilde\theta_b}$
\begin{eqnarray}
{\tilde b}_1&=& \cos {\tilde\theta_b} {\tilde b}_L 
+\sin{\tilde\theta_b} {\tilde b}_R 
\nonumber \\
{\tilde b}_2&=& -\sin{\tilde\theta_b}
 {\tilde b}_L +\cos{\tilde\theta_b} {\tilde b}_R\, .
\nonumber \\
\end{eqnarray}
At tree level, 
\begin{equation}
\sin 2 {\tilde\theta_b}
={2m_b(A_b-\mu \tan\beta)\over 
M_{{\tilde b}_1}^2 -M_{{\tilde b}_2}^2}
\label{s2bdef}
\end{equation}
and the sbottom mass eigenstates are,
\begin{equation}
M^2_{{\tilde b}_1,{\tilde b}_2}
={1\over 2}\biggl[{\tilde m}_L^2+{\tilde m}_R^2\mp
\sqrt{({\tilde m}_L^2-{\tilde m}_R^2)^2+4m_b^2 X_b^2}\biggr] \, .
\end{equation}

\subsection{$\Delta_{b}$
 Approximation:  The Effective Lagrangian Approach} 
\label{sec:db}
Loop corrections which are enhanced by powers of $\alpha_s\tan\beta$
can be included in an
effective Lagrangian approach.
At tree level, there is no $ {\overline \psi}_L b_R H_u$
 coupling in the MSSM, but such a 
coupling arises at one loop 
and gives  an effective
interaction\cite{Carena:1999py,Hall:1993gn,Guasch:2003cv}\footnote{The neutral
components of the Higgs bosons receive vacuum expectation values: $\langle
H_d^0\rangle={v_1\over \sqrt{2}},\langle
H_u^0\rangle={v_2\over \sqrt{2}}$.},
\begin{equation}
L_{eff}=-\lambda_b
{\overline \psi}_L\biggl(H_d+{\Delta_b\over \tan\beta}
H_u\biggr)b_R+h.c. \,\,\quad .
\label{effdef}
\end{equation}
Eq. \ref{effdef} shifts the $b$ quark mass from its tree level value,
\footnote{$v_{SM}=(\sqrt{2}G_F)^{-1/2}$, $v_1=v_{SM}\cos\beta$}
\begin{equation}
m_b\rightarrow{\lambda_b v_1\over \sqrt{2}} (1+\Delta_b)\, ,
\end{equation}
and  also implies that the Yukawa couplings of the
Higgs bosons to the $b$ quark are shifted from the
tree level predictions.  This shift of the
Yukawa couplings  can be included
with an effective Lagrangian approach\cite{Carena:1999py,Guasch:2003cv},
\begin{eqnarray}
L_{eff}&=&-{m_b\over v_{SM}}\biggl({1\over 1+\Delta_b}\biggr)
\biggl(-{\sin \alpha \over \cos\beta}\biggr)\biggl(1-{\Delta_b\over \tan\beta
\tan \alpha}\biggr) {\overline b} b h\, .
\label{mbdef}
\end{eqnarray}
The Lagrangian of Eq. \ref{mbdef} has been shown to
 sum all terms of ${\cal O}(\alpha_s\tan\beta)^n$ for large 
$\tan\beta$\cite{Carena:1999py,Hall:1993gn}.\footnote{It is also possible to sum the 
contributions which are proportional to $A_b$, but these terms are
less important numerically\cite{Guasch:2003cv}.}
This effective Lagrangian has
been used to compute the SQCD corrections to both the inclusive 
production process, $b {\overline b}
\rightarrow h$, and the decay process, $h\rightarrow b {\overline b}$,
and  yields results which are within a few percent of the exact
one-loop SQCD calculations\cite{Guasch:2003cv,Dittmaier:2006cz}.

The expression for $\Delta_b$ is found in the limit
 $m_b << M_h, M_Z <<M_{{\tilde b}_1}, M_{{\tilde b}_2}, M_{\tilde g}$ .  The
$1$-loop
contribution to $\Delta_b$ from sbottom/gluino loops 
is\cite{Carena:1994bv,Hall:1993gn,Carena:1999py} 
\begin{equation}
\Delta_b={2\alpha_s(\mu_S)\over 3 \pi} M_{\tilde g} \mu 
\tan\beta
I(M_{\tilde {b_1}},
M_{\tilde{ b_2}}, M_{\tilde g})\, ,
\label{db}
\end{equation}
where the function $I(a,b,c)$ is,
\begin{equation}
I(a,b,c)={1\over (a^2-b^2)(b^2-c^2)(a^2-c^2)}\biggl\{a^2b^2\log\biggl({a^2\over b^2}\biggr)
+b^2c^2\log\biggl({b^2\over c^2}\biggr)
+c^2a^2\log\biggl({c^2\over a^2}\biggr)\biggr\}\, ,
\end{equation}
and $\alpha_s(\mu_S)$ should be evaluated at a typical squark or gluino mass. 
The $2-$loop QCD corrections to $\Delta_b$ have been computed and
demonstrate that the appropriate scale at which to evaluate $\Delta_b$
is indeed of the order of the
heavy squark and gluino masses\cite{Noth:2010jy,Noth:2008tw}. The
renormalization scale 
dependence of $\Delta_b$ is minimal around $\mu_0/3$, where $\mu_0\equiv
(M_{\tilde g}+m_{\tilde {b}_1}+m_{\tilde{b}_2})/3$.  In our language this
is a high scale, of order the heavy SUSY particle masses.  The squarks
and gluinos are integrated out of the theory at this high scale and their
effects contained in $\Delta_b$.  The effective Lagrangian is then used
to calculate light Higgs production at a low scale, which is
typically the electroweak scale, $\sim 100~GeV$.

Using the effective Lagrangian of Eq. \ref{effdef}, 
which we term the Improved Born
Approximation (or
$\Delta_b$ approximation), 
the cross section is written in terms of the effective
coupling,
\begin{equation}
g_{bbh}^{\Delta_b}\equiv g_{bbh}\biggl({1\over 1+\Delta_b}\biggr)
\biggl(1-{\Delta_b\over \tan\beta
\tan \alpha}\biggr) \, ,
\label{effcoup}
\end{equation}
where
\begin{equation}
g_{bbh}=-\biggl({\sin\alpha\over \cos\beta}\biggr)
{{\overline m_b}(\mu_R)\over v_{SM}}\, .
\end{equation}
We evaluate ${\overline{m_b}}(\mu_R)$ 
using the $2-$loop ${\overline{MS}}$ value at 
a scale $\mu_R$ of ${\cal O}(M_h)$, and use the value
of $\alpha_{eff}$ determined from FeynHiggs.
The Improved Born Approximation consists of rescaling the tree level
cross section, $\sigma_0$, by the coupling of 
Eq. \ref{effcoup}\footnote{This is the approximation used in
Ref. \cite{Dittmaier:2011ti} to include the SQCD corrections.},
\begin{equation}
\sigma_{IBA}=\biggl({g_{bbh}^{\Delta_b}\over g_{bbh}}\biggr)^2\sigma_0
\, .
\label{sigibadef}
\end{equation}
The Improved Born Approximation has been shown
to accurately reproduce the full
SQCD calculation of
$pp \rightarrow {\overline t} b H^+$\cite{Berger:2003sm,Dittmaier:2009np}.

The one-loop result including the SQCD corrections
for $b g\rightarrow b h$  can be written as,
\begin{eqnarray}
\sigma_{SQCD}&\equiv&
\sigma_{IBA}\biggl(1+\Delta_{SQCD}\biggr)
\, ,
\end{eqnarray}
where $\Delta_{SQCD}$ is found from
the exact SQCD
calculation summarized in Appendix B. 

The Improved Born  Approximation involves making the replacement
in the tree level Lagrangian,
\begin{equation}
m_b\rightarrow
{m_b\over 1+\Delta_b}\, .
\label{mdef}
\end{equation}
Consistency requires that this substitution also be made in the
squark mass matrix of Eq. \ref{squark}\cite{Hofer:2009xb,Accomando:2011jy}
\begin{equation}
M_{{\tilde b}}^2\rightarrow \left(
\begin{array}{cc}
{\tilde m}_L^2 & \biggl({m_b\over 1+\Delta_b}\biggr)
 X_b\\
\biggl({m_b\over 1+\Delta_b}\biggr)
X_b & {\tilde m}_R^2\\
\end{array}
\right)\, .
\label{squark2}
\end {equation}
The effects of the substitution of Eq. \ref{mdef} in the $b$-squark
mass matrix are numerically important, although they generate contributions
which are formally higher order in $\alpha_s$.
Eqs. \ref{db} and \ref{squark2} can be solved iteratively for
$M_{{\tilde b}_1}$, $M_{{\tilde b}_2}$ and $\Delta_b$ using the
proceedure of Ref. \cite{Hofer:2009xb}\footnote{We use FeynHiggs only
for calculating $M_h$ and $\sin\alpha_{eff}$.}.

\subsection{SQCD Contributions to $g b\rightarrow b h$}

The contributions from squark and  gluino loops to the $g b\rightarrow b h$ process 
have been computed in Ref. \cite{Dawson:2007ur} in the $m_b=0$ limit.
We extend that calculation by including terms which are enhanced
by $m_b \tan\beta$ and provide analytic results in several useful limits.  

\begin{figure}[t]
\begin{center}
\includegraphics[scale=0.8]{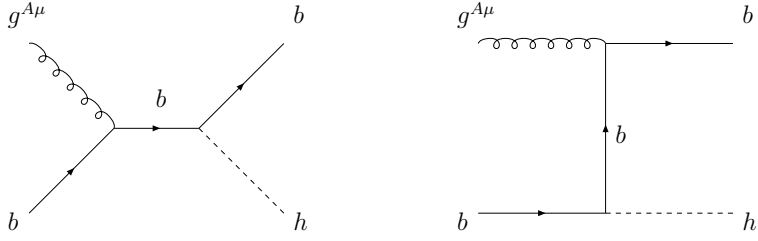} 
\caption[]{Feynman diagrams for $ g(q_1)+b (q_2)\rightarrow
b(p_b)+ h(p_h)$.}
\label{fg:bghb_feyn}
\end{center}
\end{figure}

The tree level diagrams for $g(q_1) + b(q_2) \to b(p_b) + h(p_h)$
are shown in Fig. \ref{fg:bghb_feyn}.
We define the following dimensionless spinor products 
\begin{eqnarray}
M_{s}^{\mu} & = & \frac{\overline{u}\left(p_{b}\right)\left(\slashed{q}_{1}+\slashed{q}_{2}\right)\gamma^{\mu}u\left(q_{2}\right)}{s}\nonumber \\
M_{t}^{\mu} & = & \frac{\overline{u}\left(p_{b}\right)\gamma^{\mu}\left(\slashed{p}_{b}-\slashed{q}_{1}\right)u\left(q_{2}\right)}{t}\nonumber \\
M_{1}^{\mu} & = & q_{2}^{\mu}\frac{\overline{u}\left(p_{b}\right)u\left(q_{2}\right)}{u}\nonumber \\
M_{2}^{\mu} & = & \frac{\overline{u}\left(p_{b}\right)\gamma^{\mu}u\left(q_{2}\right)}{m_{b}}\nonumber \\
M_{3}^{\mu} & = & p_{b}^{\mu}\frac{\overline{u}\left(p_{b}\right)\slashed{q}_{1}u\left(q_{2}\right)}{m_{b}t}\nonumber \\
M_{4}^{\mu} & = & q_{2}^{\mu}\frac{\overline{u}\left(p_{b}\right)\slashed{q}_{1}u\left(q_{2}\right)}{m_{b}s}\, ,
\label{eq: SME}
\end{eqnarray}
where $s=(q_1+q_2)^2, t=(p_b-q_1)^2$ and $u=(p_b-q_2)^2$.
In the  $m_b=0$ limit, the tree level amplitude depends 
only on $M_s^\mu$ and $M_t^\mu$, and $M_1^\mu$ is
generated at one-loop. 
When the effects of the $b$ mass are included, $M_2^\mu$, $M_3^\mu$, and $M_4^\mu$ 
are also generated.

The tree level amplitude is 
\begin{eqnarray}
\mathcal{A}_{\alpha\beta}^{a}\mid_0 & = & -g_{s}
g_{bbh}\left(T^{a}\right)_{\alpha\beta}\epsilon_{\mu}(q_1)\left\{ M_{s}^{\mu}+M_{t}^{\mu}\right\} \, ,
\end{eqnarray}
and the one loop contribution can be written as 
\begin{equation}
\mathcal{A}_{\alpha\beta}^{a}
=-\frac{\alpha_{s}(\mu_R)}{4\pi}g_{s}g_{bbh}\left(T^{a}\right)_{\alpha\beta}\sum_{j}X_{j}M_{j}^{\mu}\epsilon_{\mu}(q_1)\, .
\label{onedef}
\end{equation}
In the calculations to follow, only the non-zero $X_j$ coefficients
 are listed and we neglect terms of ${\cal O}(m_b^2/s)$ if they are
not enhanced by $\tan\beta$.

The renormalization of the 
squark and gluino contributions 
is performed in the on-shell scheme and has been described 
in Refs. \cite{Dawson:2007ur,Berge:2007dz,Noth:2010jy}.
The bottom quark self-energy is
\begin{eqnarray}
\Sigma_{b}\left(p\right) & = & \slashed{p}
\biggl(\Sigma_{b}^{V}(p^2)-\Sigma_{b}^{A}(p^2)
\gamma_{5}\biggr)+m_{b}\Sigma_{b}^{S}(p^2)\, .
\end{eqnarray}
The $b$ quark fields are renormalized as $b\rightarrow \sqrt{Z_b^V}
b$ and $Z_b^V\equiv \sqrt{1+\delta Z_b^V}$.
The contribution from the counter-terms to the self-energy is, 
\begin{eqnarray}
\Sigma_{b}^{\mathrm{ren}}\left(p\right) 
& = & \Sigma_{b}\left(p\right)+\delta\Sigma_b(p)\nonumber \\
\delta\Sigma_{b}\left(p\right)&=&\slashed{p}\left(\delta Z_{b}^{V}-\delta Z_{b}^{A}\gamma_{5}\right)-m_{b}\delta Z_{b}^{V}-\delta m_{b}\, .
\end{eqnarray}
Neglecting the $\gamma_5$ contribution,
the renormalized self-energy is then given by 
\begin{eqnarray}
\Sigma_{b}^{\mathrm{ren}}\left(p\right) 
 & = & \left(\slashed{p}-m_{b}\right)
\left(\Sigma_{b}^{V}(p^2)+\delta Z_{b}^{V}\right)+m_{b}\left(
\Sigma_{b}^{S}(p^2)+\Sigma_{b}^{V}(p^2)-\frac{\delta m_{b}}{m_{b}}\right)\, .
\end{eqnarray}
The on-shell renormalization condition implies
\begin{eqnarray}
\left.\Sigma_{b}^{\mathrm{ren}}\left(p\right)\right|_{\slashed{p}=m_{b}} & = & 0\\
 lim_{\slashed{p}\rightarrow m_b}
\biggl(
\frac{\Sigma_{b}^{\mathrm{ren}}\left(p\right)}{\slashed{p}-m_{b}}
\biggr)
& = & 0
\, .
\end{eqnarray}

The mass and wavefunction counter-terms are\footnote{$s_{2\tilde{b}}\equiv
\sin 2{\tilde \theta}_b$.}
\begin{eqnarray}
\frac{\delta m_{b}}{m_{b}} & = & \left[\Sigma_{b}^{S}\left(p^{2}\right)+\Sigma_{b}^{V}\left(p^{2}\right)\right]_{p^{2}=m_{b}^{2}}\nonumber \\
 & = & \frac{\alpha_{s}(\mu_R)}{3\pi}\sum_{i=1}^{2}\left[\left(-1\right)^{i}\frac{M_{\tilde{g}}}{m_{b}}s_{2\tilde{b}}B_{0}-B_{1}\right]\left(0;M_{\tilde{g}}^{2},M_{\tilde{b}_{i}}^{2}\right)\label{eq: del mb}\\
\delta Z_{b}^{V} & = & -\left.\Sigma_{b}^{V}\left(p^{2}\right)\right|_{p^{2}=m_{b}^{2}}
-2m_b^2
{\partial\over \partial p^2} \biggl(\Sigma_b^V(p^2)
+\Sigma_S(p^2)\biggr)\mid_{p^2=m_b^2}\nonumber \\
 & = & \frac{\alpha_{s}(\mu_R)}{3\pi}\sum_{i=1}^{2}\biggl[B_{1}
+2m_b^2 B_1^\prime-(-1)^i 2m_b M_{\tilde g} s_{2{\tilde b}}B_0^\prime
\biggr]
\left(0;M_{\tilde{g}}^{2},M_{\tilde{b}_{i}}^{2}\right)
\, ,\label{eq:delZ}
\end{eqnarray}
where we consistently neglect the $b$ quark mass if it is not enhanced by $\tan\beta$.
The
Passarino-Veltman functions $B_0\left(0;M_{\tilde{g}}^{2},M_{\tilde{b}_{i}}^{2}\right)$
and $B_1\left(0;M_{\tilde{g}}^{2},M_{\tilde{b}_{i}}^{2}\right)$ are defined in Appendix A.  
Using the tree level relationship of Eq. \ref{s2bdef}, the mass counterterm
can be written as,
\begin{eqnarray}
{\delta m_b\over m_b} 
&=& 
{2\alpha_s(\mu_R)\over 3\pi}
M_{\tilde g} A_b 
I(M_{{\tilde b}_1},M_{{\tilde b}_2}, M_{\tilde g})
-\Delta_b
-
\frac{\alpha_{s}(\mu_R)}{3\pi}\sum_{i=1}^{2}
 B_1\left(0;
M_{\tilde{g}}^{2},
M_{\tilde{b}_{i}}^{2}\right)
\, .
\label{mbnice}
\end{eqnarray}

The external gluon is renormalized as 
$g_\mu^A\rightarrow 
\sqrt{Z_3}g_\mu^A=
\sqrt{1+\delta Z_3}g_\mu^A$
 and the strong coupling renormalization is $g_s\rightarrow
Z_g g_s$ with $\delta Z_g=-\delta Z_3/2$.  We renormalize $g_s$ using
the ${\overline {MS}}$ scheme with the 
heavy squark and gluino contributions
subtracted at zero momentum\cite{Nason:1987xz},
\begin{equation}
\delta Z_3=-
{\alpha_s(\mu_R)\over 4 \pi}
\biggl[
{1\over 6}
\Sigma_{{\tilde q}_i}
\biggl(
{4\pi\mu_R^2\over M_{{\tilde q}_i}^2}
\biggr)^\epsilon
+2\biggl(
{4\pi\mu_R^2\over M_{\tilde g}^2}
\biggr)^\epsilon
\biggr]
{1\over \epsilon}\Gamma(1+\epsilon)\, .
\end{equation}

In order to avoid overcounting the effects which are contained in 
$g_{bbh}^{\Delta_b}$ to ${\cal O} (\alpha_s)$, we need the additional
counterterm,
\begin{equation}
\delta_{CT}=\Delta_b\biggl( 1+{1\over \tan\beta\tan\alpha}\biggr)
\, .
\label{ctdef}
\end{equation}
The total contribution of the counterterms is, 
\begin{equation}
\sigma_{CT}=\sigma_{IBA}\biggl(
2 \delta Z_b^V+\delta Z_3+2 \delta Z_g+2{\delta m_b\over m_b}+2
\delta_{CT}\biggr)
=2\sigma_{IBA}\biggl(
\delta Z_b^V+{\delta m_b\over m_b}+
\delta_{CT}\biggr)
\, .
\label{cttot}
\end{equation}
The $\tan\beta$ enhanced contributions from 
$\Delta_b$ cancel between Eqs. \ref{mbnice} and \ref{ctdef}.
The expressions for the contributions to the $X_i$, 
as defined in Eq. \ref{onedef},
are given in Appendix B for arbitrary squark and gluino masses,
 and separately for each $1-$ loop diagram. 

\section{Results for Maximal and Minimal Mixing in the $b$-Squark Sector}

\subsection{Maximal Mixing}

The squark and gluino contributions to $bg\rightarrow bh$
can be examined analytically in several
scenarios.  In the first scenario, 
\begin{equation}
\mid {\tilde m}_L^2
-{\tilde m}_R^2\mid << {m_b\over 1+\Delta_b}
\mid X_b\mid\, .
\end{equation} 
 We expand in powers
of   ${\mid {\tilde m}_L^2
-{\tilde m}_R^2\mid \over m_b X_b}$.  In this case the sbottom masses
are nearly degenerate,
\begin{eqnarray}
 M_S^2&\equiv &{1\over 2}
\biggl[
M_{{\tilde b}_1}^2+M_{{\tilde b}_2}^2
\biggr]
\nonumber \\
\mid M_{{\tilde b}_1}^2-M_{{\tilde b}_2}^2\mid
&=&
\biggl({2m_b \mid X_b\mid \over 1+\Delta_b}\biggr)
\biggl(1+{( {\tilde m}_L^2
-{\tilde m}_R^2)^2 (1+\Delta_b)^2\over 8  m_b^2 X_b^2}\biggr)<< M_S^2
\, .
\end{eqnarray}
This scenario is termed maximal mixing since
\begin{equation}
\sin 2{{\tilde{\theta}}}_b\sim 1-
{({\tilde m}_L^2
-{\tilde m}_R^2)^2(1+\Delta_b)^2\over 8  m_b^2 X_b^2}
\, .
\end{equation}
We expand the contributions of
the exact one-loop SQCD calculation 
given in Appendix B in powers of $1/M_S$, keeping terms
to ${\cal O}\biggl({M_{EW}^2\over M_S^2}\biggr)$ and assuming
$M_S\sim M_{\tilde g}\sim \mu\sim A_b\sim
{\tilde m}_L\sim {\tilde m}_R >> M_W, M_Z, M_h\sim M_{EW}$.
In the expansions, we assume the
large $\tan\beta$ limit and take 
$m_b\tan\beta\sim {\cal {O}}(M_{EW})$.
This expansion has been studied in
detail for the decay 
$h\rightarrow b {\overline{b}}$, 
with particular emphasis
on the decoupling properties of the results as 
$M_S$ and $M_{\tilde g}
\rightarrow\infty$\cite{Haber:2000kq}.
The SQCD contributions to the decay, $h\rightarrow b {\overline b}$,
extracted from our results are in agreement with those of Refs. 
\cite{Haber:2000kq,Accomando:2011jy}

The final result for maximal mixing, summing all contributions, is,
\begin{eqnarray}
A_s &\equiv &-g_s T^A g_{bbh}M_s^\mu \biggl\{
1+{\alpha_s(\mu_R)\over 4 \pi }
X_i^s\biggr\}
\nonumber \\
&=&-g_s T^A g_{bbh}M_s^\mu\biggl\{
1+\biggl({\delta g_{bbh}\over g_{bbh}}\biggr)_{max}
+{\alpha_s(\mu_R)\over 4\pi}
{s\over M_S^2}\delta \kappa_{max}\biggr\}
\nonumber \\
A_t
&\equiv &-g_s T^A g_{bbh}M_s^\mu \biggl\{
1+{\alpha_s(\mu_R)\over 4 \pi }
X_i^t\biggr\}
\nonumber \\
&=&-g_s T^A g_{bbh} M_t^\mu\biggl\{
1+\biggl({\delta g_{bbh}\over g_{bbh}}\biggr)_{max}\biggr\}
\nonumber \\
A_1
&\equiv &-g_s T^A g_{bbh}M_s^\mu \biggl\{
1+{\alpha_s(\mu_R)\over 4 \pi }
X_i^1\biggr\}
\nonumber \\
&=&-g_s T^A g_{bbh}M_1^\mu \biggl(
-{\alpha_s(\mu_R) u\over 2 \pi M_S^2}
\biggr)
\delta \kappa_{max}\, .
\label{ansmax}
\end{eqnarray}

The contribution which is a rescaling of the $b {\overline b} h$ vertex is,
\begin{equation}
\biggl({\delta g_{bbh}\over g_{bbh}}\biggr)_{max}=
\biggl({\delta g_{bbh}\over g_{bbh}}\biggr)^{(1)}_{max}
+\biggl({\delta g_{bbh}\over g_{bbh}}\biggr)^{(2)}_{max}\, ,
\end{equation}
where the leading order term in $M_{EW}/M_S$ is ${\cal O}(1)$,
\begin{equation}
\biggl({\delta g_{bbh}\over g_{bbh}}\biggr)^{(1)}_{max}
={\alpha_s(\mu_R)\over 3 \pi}
{M_{\tilde g} (X_b-Y_b)\over M_S^2}
f_1(R)\, ,
\label{lomax}
\end{equation}
with $Y_b\equiv A_b+\mu\cot\alpha$ and $R\equiv M_{\tilde g}/M_S$.
Eq. \ref{lomax} only decouples for large $M_S$ if the additional
limit $M_A\rightarrow \infty$ is also 
taken\cite{Haber:2000kq,Dawson:2007ur}.  In this limit,
\begin{equation}
X_b-Y_b\rightarrow {2\mu M_Z^2\over M_A^2}
\tan\beta\cos 2\beta + {\cal O}\biggl({M_{EW}^4\over M_A^4}
\biggr)\, .
\end{equation}

The subleading terms of ${\cal O}(M_{EW}^2/M_S^2)$
are,\footnote{We use the shorthand, $c_\beta=
\cos\beta$, $s_{\alpha+\beta}=\sin(\alpha+\beta)$, etc.}
\begin{eqnarray}
\biggl({\delta g_{bbh}\over g_{bbh}}\biggr)^{(2)}_{max}
&=&{\alpha_s(\mu_R)\over 3 \pi}
\biggl\{
-{M_{\tilde g} Y_b\over M_S^2}
\biggl[{M_h^2\over 12 M_S^2} f_3^{-1}(R)
+{X_b^2 m_b^2\over 2 (1+\Delta_b)^2 M_S^4}f_3(R)\biggr]
\nonumber \\ &&
-{m_b^2 X_bY_b\over 2 (1+\Delta_b)^2 M_S^4}f_3^{-1}(R)\nonumber \\
&&+
{M_Z^2\over 3 M_S^2}
{c_\beta s_{\alpha+\beta}\over s_\alpha}
I_3^b\biggl[
3f_1(R)+\biggl(
{2 M_{\tilde g} X_b\over M_S^2}-1\biggr)f_2(R)\biggr]\biggr\}\,
\label{dg2max}
\end{eqnarray}
The functions $f_i(R)$ are
defined in Appendix C.

\begin{figure}[t]
\begin{center}
\includegraphics[scale=0.6]{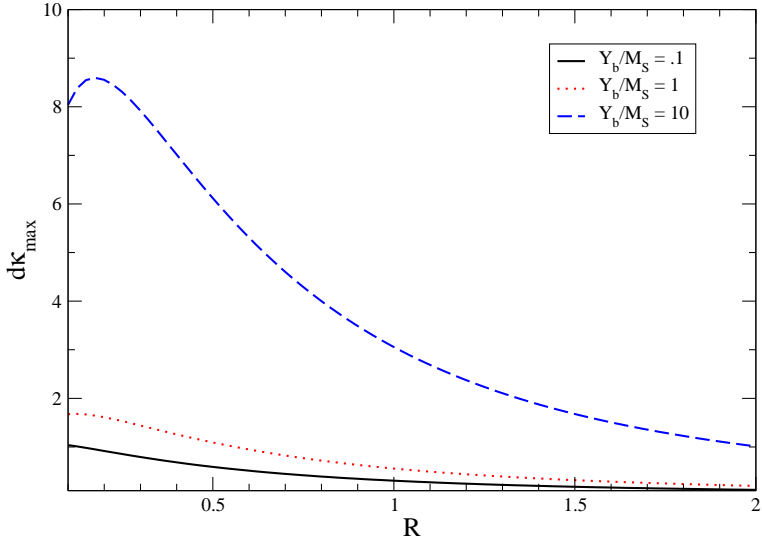} 
\caption[]{Contribution of $\delta \kappa_{max}$ defined in Eq. 
\ref{dkapmax} as a function of $R=M_{\tilde g}/M_S$.}
\label{fg:dk_max}
\end{center}
\end{figure}

 The ${s\over M_S^2},{u\over M_S^2}$ terms in Eq. \ref{ansmax} 
 are not a rescaling of the lowest order vertex and
cannot be obtained from the effective Lagrangian.  We find,
\begin{equation}
\delta \kappa_{max}=
{1\over 4}
\biggl[
f_3(R)+{1\over 9}f_3^{-1}(R)
\biggr]
-R{Y_b\over 2 M_S}\biggl[
f_2^\prime(R)+{1\over 9}{\hat f}_2(R)\biggr]\, .
\label{dkapmax}
\end{equation}
The $\delta\kappa_{max}$ term
is ${\cal O}(1)$ in $M_{EW}/M_S$ and has its largest values
 for small $R$ and large ratios of $Y_b/M_S$, as can be 
seen in Fig. \ref{fg:dk_max}. 
Large effects can be obtained for $Y_b/M_S \sim 10$ and
$M_{\tilde g} <<  M_S$.   
However, the parameters must be carefully tuned so that
$A_b/M_S\lsim 1$ in order not to break color\cite{Gunion:1987qv}.

 The amplitude squared, summing over final state spins and colors and averaging
 over initial state spins and colors, including one-loop SQCD
 corrections is 
\begin{eqnarray}
\left|
{\overline{\mathcal{A}}}\right|_{max}^{2} 
 & = & -\frac{2\pi\alpha_{s}(\mu_R)}{3}g_{bbh}^{2}\left[\left(
\frac{u^{2}+M_{h}^{4}}{st}\right)\left[1+2
\biggl({\delta g_{bbh}\over g_{bbh}}\biggr)_{max}
\right]+{\alpha_s(\mu_R)\over 2\pi}
\frac{M_{h}^{2}}{M_{S}^{2}}\delta\kappa_{max}\right]\, .
\label{eq:amp_sq_del}
\end{eqnarray}
Note that in the cross section, the $\delta \kappa_{max}$ term is not enhanced
by a power of $s$ and gives a contribution of 
${\cal O}\biggl({M_{EW}^2\over 
M_S^2}\biggr)$.

Expanding $\Delta_b$ in the maximal mixing limit,
\begin{equation}
\Delta_b\rightarrow -{\alpha_s (\mu_S)\over 3 \pi}
{M_{\tilde g}\mu\over M_S^2}\tan\beta f_1(R)+{\cal O}\biggl(
{M_{EW}^4\over M_S^4}\biggr)\, .
\label{mbmaxlim}
\end{equation}
By comparison with Eq. \ref{effcoup},
\begin{eqnarray}
\left|
{\overline{\mathcal{A}}}\right|_{max}^{2} 
 & = & -\frac{2\pi\alpha_{s}(\mu_R)}{3}(g_{bbh}^{\Delta_b})^{2}
\left\{\left(
\frac{u^{2}+M_{h}^{4}}{st}\right)\left[1+2
\biggl({\delta g_{bbh}\over g_{bbh}}\biggr)^{(2)}_{max}
\right]\right.
\nonumber \\ &&
\left.+{\alpha_s(\mu_R)\over 2\pi}
\frac{M_{h}^{2}}{M_{S}^{2}}\delta\kappa_{max}\right\}
+{\cal O}\biggl(\biggl[{M_{EW}\over M_S}\biggr]^4,\alpha_s^3\biggr)\, .
\label{maxans}
\end{eqnarray}
Note that the mis-match in the arguments of $\alpha_s$ in Eqs.
\ref{mbmaxlim} and \ref{maxans} is higher order in $\alpha_s$ than
the terms considered here.
The $(\delta g_{bbh}/g_{bbh})^{(2)}_{max}$ and 
$\delta\kappa_{max}$ terms both correspond to contributions which are
not present in the effective Lagrangian approach.  These terms are, however,
suppressed by powers of $M_{EW}^2/M_S^2$
and   the
non-decoupling effects discussed in Refs. \cite{Haber:2000kq}
and \cite{Guasch:2003cv} are completely contained
in the $g_{bbh}^{\Delta_b}$ term.

\subsection{Minimal Mixing in the $b$ Squark Sector}
The minimal mixing scenario is characterized by a mass splitting between the $b$ squarks
which is of order the $b$ squark mass,
$\mid M_{\tilde{b}_1}^2-M_{\tilde{b}_2}^2\mid \sim M_S^2$.  
In this case, 
\begin{equation}
\mid {\tilde m}_L^2-{\tilde m}_R^2\mid >> {m_b
\mid X_b\mid\over (1+\Delta_b)}\, ,
\end{equation} 
and the mixing angle in the $b$ squark sector is close to zero,
\begin{equation}
\cos 2 {\tilde \theta}_b\sim 1-{2m_b^2X_b^2\over 
(M_{{\tilde b}_1}^2-M_{{\tilde b}_2}^2)^2}
\biggl({1\over 1+\Delta_b}\biggr)^2
\, .
\end{equation}
 
\begin{figure}[t]
\begin{center}
\includegraphics[scale=0.6]{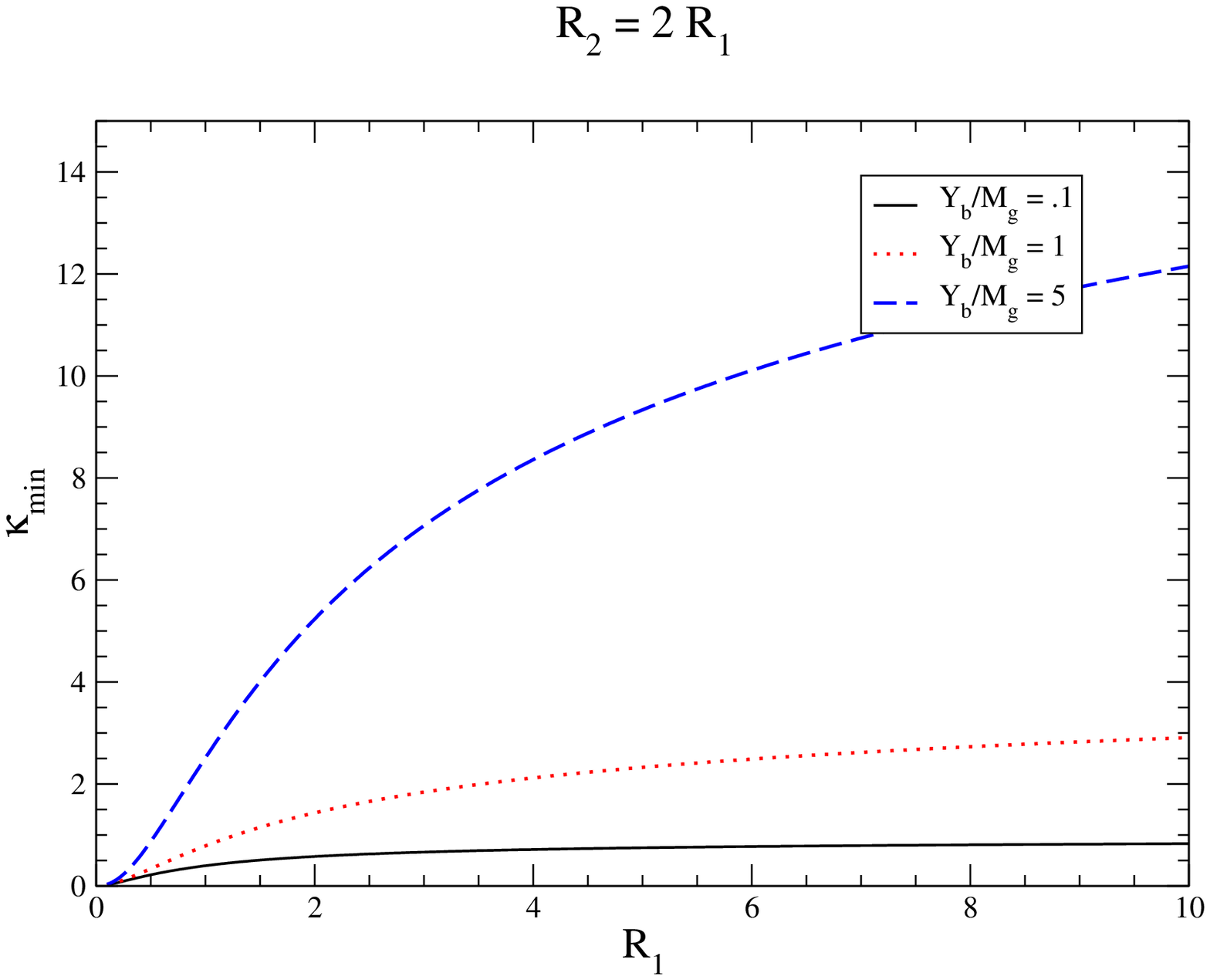} 
\caption[]{Contribution of $\delta \kappa_{min}$ defined
in Eq. \ref{minkapdef} as a function of 
$R_i=M_{\tilde g}/M_{{\tilde b}_i}$
.}
\label{fg:dk_min}
\end{center}
\end{figure}

The non-zero subamplitudes are
\begin{eqnarray}
A_s &=&-
g_s T^A g_{bbh}M_s^\mu\biggl\{
1+\biggl({\delta g_{bbh}\over g_{bbh}}\biggr)_{min}
+{\alpha_s(\mu_R)\over 4\pi}
{s\over {\tilde M}_g^2}\delta \kappa_{min} \biggr\}\nonumber \\
A_t&=&-g_s T^A g_{bbh} M_t^\mu\biggl\{
1+\biggl({\delta g_{bbh}\over g_{bbh}}\biggr)_{min}\biggr\}
\nonumber \\
A_1&=&-g_s T^A g_{bbh}M_1^\mu \biggl(-
{\alpha_s(\mu_R) u\over 2 \pi {\tilde M}_g^2}
\biggr)
\delta \kappa_{min}
\, .
\end{eqnarray}
Expanding the exact one-loop results of Appendix B in the minimal mixing
scenario,  
\begin{equation}
\delta \kappa_{min}={ 1\over 8}\Sigma_{i=1}^2
\biggl(R_i^2\biggl[{1\over 9}f_3^{-1}(R_i)
+f_3(R_i)\biggr]\biggr)
+{ Y_b\over M_{\tilde g}}
{R_1^2 R_2^2\over R_2^2-R_1^2}
\biggl(
3h_1(R_1,R_2,1)+{8\over 3} h_1(R_1,R_2,2)\biggr)
\, ,
\label{minkapdef}
\end{equation}
where $R_i=M_{\tilde g}/M_{\tilde{b}_i}$ 
and the functions $f_i(R_i)$ and $h_i(R_1,R_2,n)$
are defined in Appendix C. The $\delta \kappa_{min}$
function is shown in Fig. \ref{fg:dk_min}.  For large
values of $Y_b/ M_{\tilde g}$ it can be significantly larger
than $1$.

As in the previous section, the  spin and color 
averaged amplitude-squared is,
\begin{eqnarray}
\mid {\overline A}\mid_{min}^2 
&=&-{2\alpha_s(\mu_R)\pi\over 3}(g_{bbh}^2)\biggl\{
{(M_h^4+u^2)\over s t}\biggl[1+2\biggl({\delta g_{bbh}\over
g_{bbh}}\biggr)_{min}\biggr]+{\alpha_s(\mu_R)
\over 2\pi}\delta \kappa_{min}
{M_h^2\over M_{\tilde g}^2}
\biggr\}
\, ,
\end{eqnarray}
with,
\begin{equation}
\biggl({\delta g_{bbh}\over g_{bbh}}\biggr)_{min}=
\biggl({\delta g_{bbh}\over g_{bbh}}\biggr)^{(1)}_{min}
+\biggl({\delta g_{bbh}\over g_{bbh}}\biggr)^{(2)}_{min}
\, .
\end{equation}
The leading order term in $M_{EW}/M_S$ is ${\cal O}(1)$,
\begin{equation}
\biggl(
{\delta g_{bbh}\over g_{bbh}}
\biggr)^{(1)}_{min}
={2\alpha_s(\mu_R)\over 3 \pi}
{ (X_b-Y_b)\over M_{\tilde g}}{R_1^2 R_2^2\over
R_1^2-R_2^2} h_1(R_1,R_2,0)\, .
\label{mindg1}
\end{equation}

The subleading terms are ${\cal O}\biggl({M_{EW}^2\over M_S^2}\biggr)$,
\begin{eqnarray}
\biggl({\delta g_{bbh}\over g_{bbh}}\biggr)^{(2)}_{min}
&=&{\alpha_s\over 4\pi}
\biggl\{
 -\frac{8M_{\tilde{g}}Y_{b}}{3\Delta M_{\tilde{b}_{12}}^{2}}
\left[
\frac{h_{2}\left(R_{1},R_{2}\right)M_{h}^{2}}
{\Delta M_{\tilde{b}_{12}}^{2}}
\right.
\nonumber \\ &&
\left. \left. +\frac{m_{b}^{2}X_b^{2}}
{\left(\Delta M_{\tilde{b}_{12}}^{2}\right)^{2}
(1+\Delta_b)^2}
\left\{ 
2\mathcal{S}\left(\frac{f_{1}\left(R\right)}
{M_{\tilde{b}}^{2}}\right)\right.\right.\left.
+\frac{h_{1}\left(R_{1},R_{2},0\right)}
{\Delta M_{\tilde{b}_{12}}^{2}}
\right\} 
\right]
\nonumber \\ &&
+\frac{4}{3}\frac{c_{\beta}s_{\alpha+\beta}}
{s_{\alpha}}I_{3}^{b}M_{Z}^{2}
\left[\mathcal{S}\left(\frac{3f_{1}\left(R\right)
-f_{2}\left(R\right)}{3M_{\tilde{b}}^{2}}\right)
-\frac{2M_{\tilde{g}}X_b}
{\Delta M_{\tilde{b}_{12}}^{2}}
\mathcal{A}\left(\frac{f_{1}\left(R\right)}
{M_{\tilde{b}}^{2}}\right)\right]
\nonumber \\ &&
 +\frac{4}{3}
\frac{c_{\beta}s_{\alpha+\beta}}
{s_{\alpha}}\left(I_{3}^{b}
-2Q^{b}s_{W}^{2}\right)
M_{Z}^{2}\left[\mathcal{A}
\left(\frac{3f_{1}\left(R\right)-f_{2}\left(R\right)}
{3M_{\tilde{b}}^{2}}\right)\right.
\nonumber \\ &&
\left.
 -\frac{2M_{\tilde{g}}X_b}
{\Delta M_{\tilde{b}_{12}}^{2}}
\left\{ \mathcal{S}\left(\frac{f_{1}\left(R\right)}
{M_{\tilde{b}}^{2}}\right)
+\frac{h_{1}\left(R_{1},R_{2},0\right)}
{\Delta M_{\tilde{b}_{12}}^{2}}\right\} \right]\nonumber \\
 &  & +
\frac{8}{3}\frac{m_{b}^{2}X_{b}Y_{b}}
{\Delta M_{\tilde{b}_{12}}^{2}(1+\Delta_b)^2}\mathcal{A}\left(\frac{3f_{1}\left(R\right)-f_{2}\left(R\right)}{3M_{\tilde{b}}^{2}}\right)\biggr\}
\, .
\label{dg2min}
\end{eqnarray}
The symmetric and anti-symmetric
functions are defined,
\begin{eqnarray}
{\mathcal{S}}(f(R,M_{\tilde b})
&\equiv & {1\over 2} \biggl[ f(R_1, M_{{\tilde b}_1})+
f(R_2, M_{{\tilde b}_2})\biggr]
\nonumber \\
{\mathcal{A}}(f(R,M_{\tilde b})
&\equiv & {1\over 2} \biggl[ f(R_1, M_{{\tilde b}_1})-
f(R_2, M_{{\tilde b}_2})\biggr]
\end{eqnarray}
and $\Delta M^2_{{\tilde b}_{12}}\equiv M_{\tilde{b}_1}^2-
 M_{\tilde{b}_2}^2$.  The remaining functions are defined 
in Appendix C.

By expanding $\Delta_b$ in the minimal mixing limit, we find
the analogous result to that of the maximal mixing case,
\begin{eqnarray}
\mid {\overline A}\mid_{min}^2 
&=&-{2\alpha_s\pi\over 3}(g_{bbh}^{\Delta_b})^2\biggl\{
{(M_h^4+u^2)\over s t}\biggl[1+2\biggl({\delta g_{bbh}\over
g_{bbh}}\biggr)^{(2)}_{min}\biggr]
\nonumber \\ && 
+{\alpha_s\over 2\pi}\delta \kappa_{min}
{M_h^2\over M_{\tilde g}^2}
\biggr\}+{\cal O}\biggl(\biggl[{M_{EW}\over M_S}\biggr]^4,\alpha_s^3\biggr)
\, .
\label{minansdef}
\end{eqnarray}
The contributions which are not contained in $\sigma_{IBA}$ are
again found to be
suppressed by ${\cal O}\biggl(\biggl[{M_{EW}\over M_S}\biggr]^2\biggr)$.

\section{Numerical Results}

\begin{figure}[t]
\begin{center}
\includegraphics[scale=0.6]{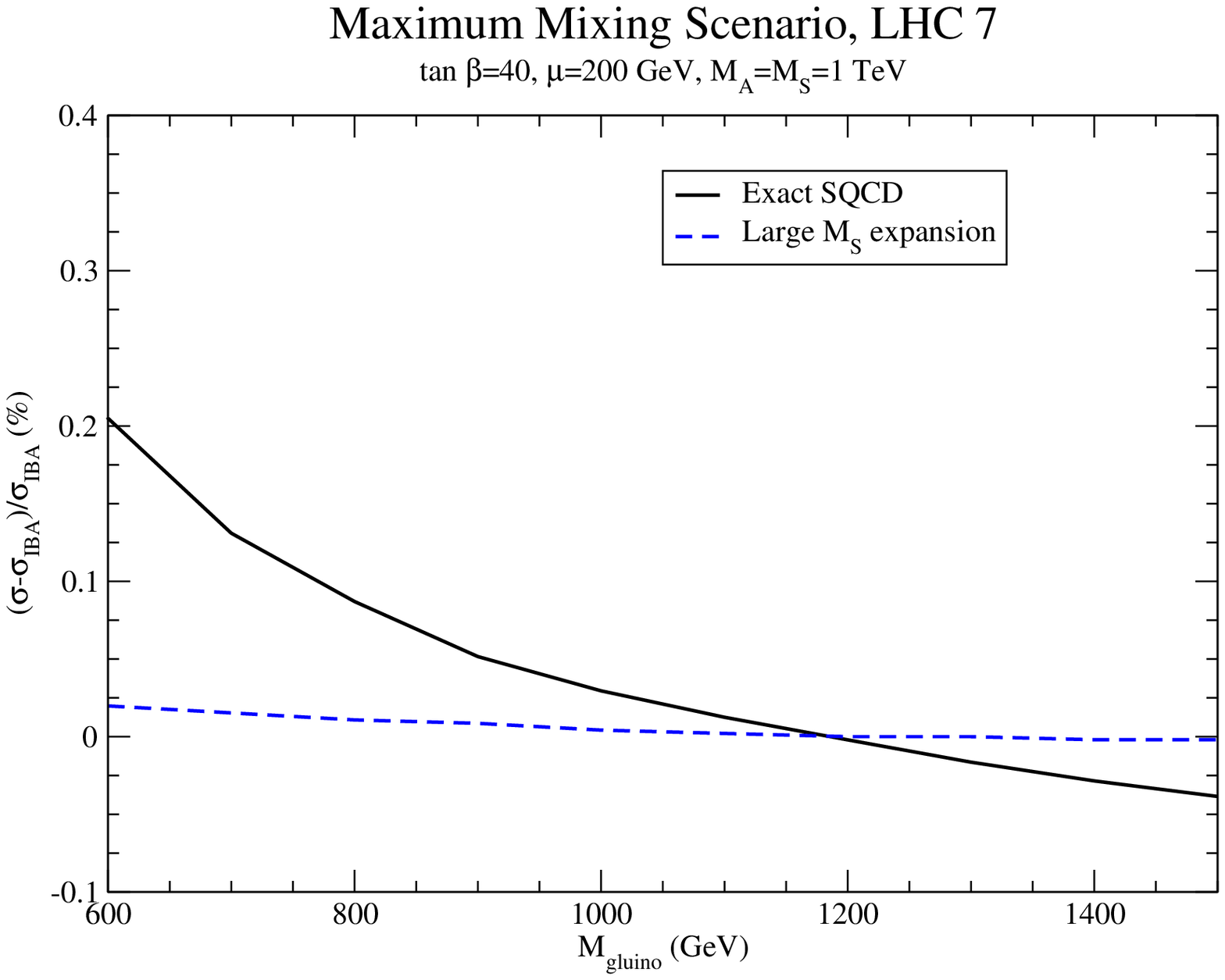} 
\caption[]{Percentage difference between the Improved Born Approximation and
the exact one-loop SQCD calculation of $pp\rightarrow
bh$ for maximal mixing in the $b$-squark 
sector at $\sqrt{s}=7~TeV$, $\tan\beta=40$, and $M_A=1~TeV$.}
\label{fg:maxmix}
\end{center}
\end{figure}

\begin{figure}[t]
\begin{center}
\includegraphics[scale=0.6]{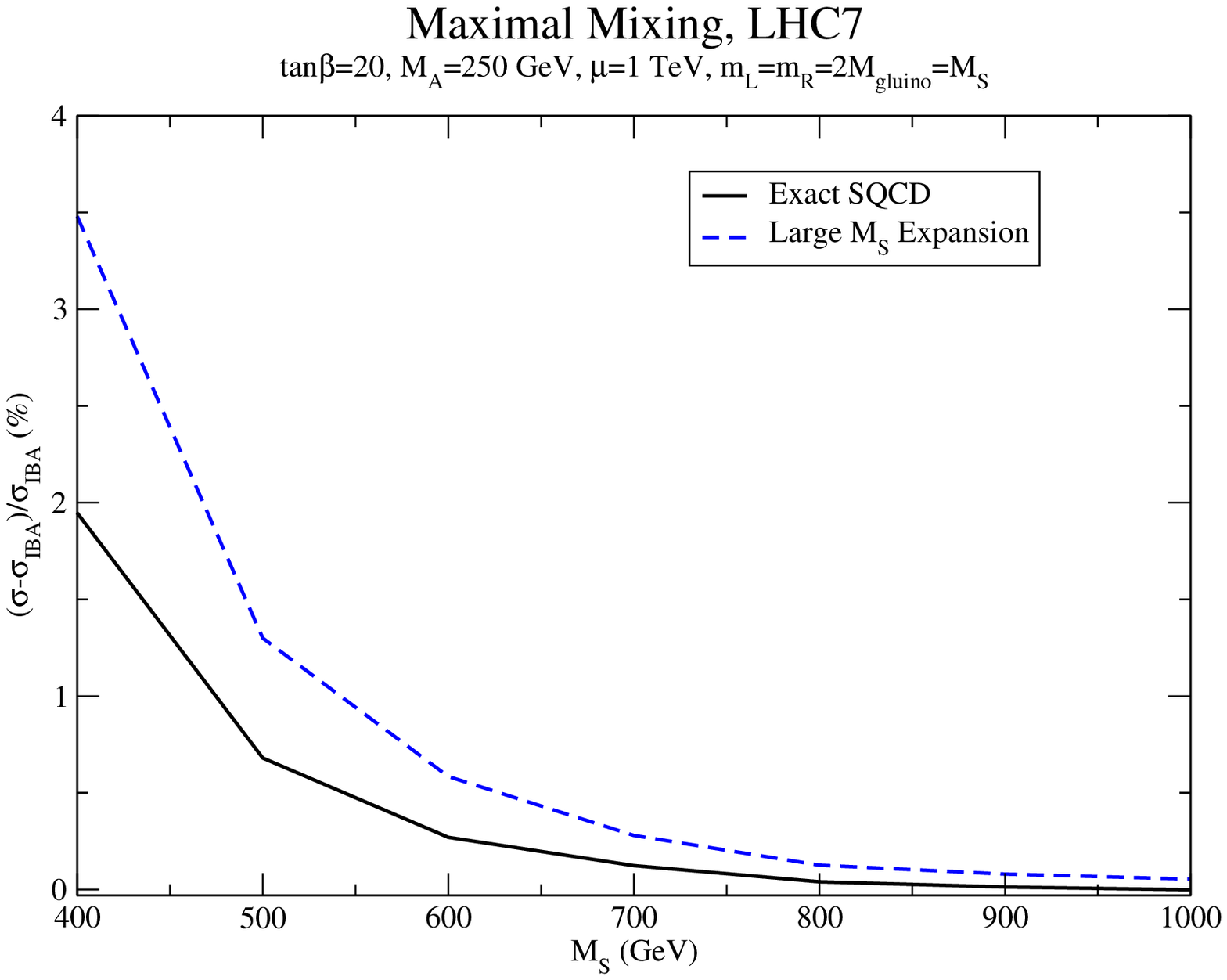} 
\caption[]{Percentage difference between the Improved Born Approximation and
the exact one-loop SQCD calculation of $pp\rightarrow
bh$ for maximal mixing in the $b$-squark 
sector at $\sqrt{s}=7~TeV$, $\tan\beta=20$, and $M_A=250~GeV$.}
\label{fg:maxmix250}
\end{center}
\end{figure}

\begin{figure}[t]
\begin{center}
\includegraphics[scale=0.6]{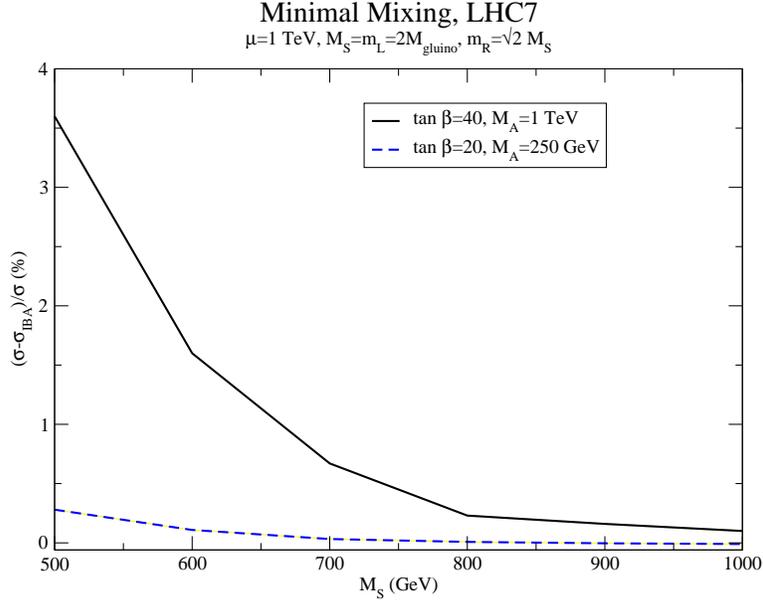} 
\caption[]{Percentage difference between the Improved Born Approximation
and the exact one-loop SQCD calculation for $pp
\rightarrow bh$ for minimal mixing in the
$b$ squark sector at $\sqrt{s}=7~ TeV$.}
\label{fg:minmix}
\end{center}
\end{figure}

\begin{figure}[t]
\begin{center}
\includegraphics[scale=0.6]{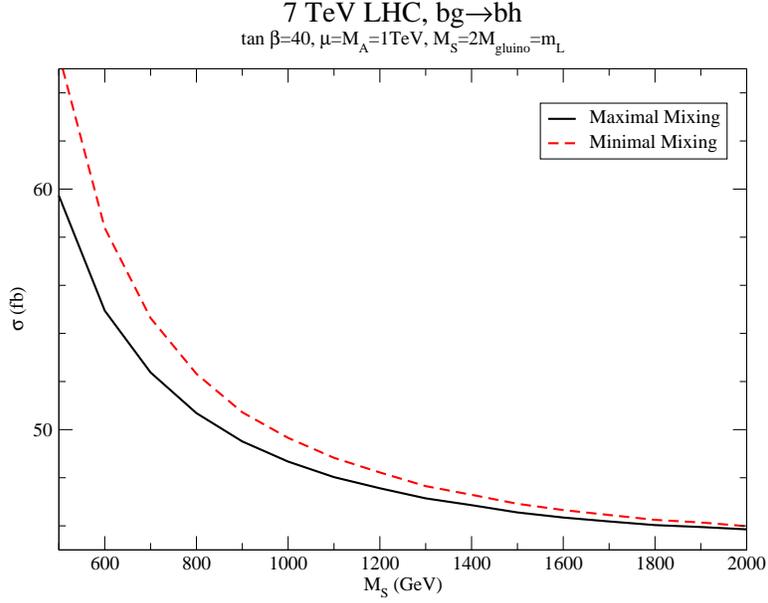} 
\caption[]{Comparison between the exact one-loop SQCD calculation for $pp
\rightarrow bh$ for minimal and maximal mixing in the
$b$ squark sector at $\sqrt{s}=7~ TeV$ and $\tan\beta=40$.
The minimal mixing curve has $m_R=\sqrt{2}M_S$ and 
${\tilde{\theta}}_b\sim 0$, while
the maximal mixing curve has $m_R=M_S$ and 
${\tilde{\theta}}_b\sim {\pi\over 4}$. }
\label{fg:compsig1}
\end{center}
\end{figure}

We present results for $pp\rightarrow b ({\overline b})h$ at $\sqrt{s}
= 7~TeV$ with $p_{Tb}>20~GeV$ and $\mid \eta_b\mid < 2.0$.   We use
FeynHiggs to generate $M_h$ and $\sin\alpha_{eff}$ and then iteratively
solve for the $b$ squark masses and $\Delta_b$ from Eqs. \ref{db}
and \ref{squark2}.  We evaluate the 2-loop ${\overline {MS}}$ $b$
mass at $\mu_R=M_h/2$, which we also take to be the renormalization and
factorization 
scales\footnote{$\Delta_b$ is evaluated using $\alpha_s(M_S)$.}.
Finally,   Figs \ref{fg:maxmix}, \ref{fg:maxmix250},
 \ref{fg:minmix}, and \ref{fg:compsig1}
use
the CTEQ6m NLO parton distribution functions\cite{Nadolsky:2008zw}.
 Figs. 
\ref{fg:maxmix}, \ref{fg:maxmix250}
 and \ref{fg:minmix} show the percentage deviation of
 the complete one-loop
SQCD calculation from the Improved Born Approximation of Eq. 
\ref{sigibadef} for
$\tan\beta=40$ and $\tan\beta=20$ and representative values of the MSSM 
parameters\footnote{Figs. \ref{fg:maxmix}, \ref{fg:maxmix250} and \ref{fg:minmix}
do not include the pure QCD NLO corrections\cite{Dicus:1998hs}.}.  
In both
extremes of $b$ squark mixing, the Improved Born Approximation
 approximation is within a
few percent of the complete one-loop
SQCD calculation and so is a reliable prediction for
the rate. This is true for both large and small $M_A$.
 In addition, the large $M_S$ expansion accurately 
reproduces the full SQCD one-loop result to within a few percent.
 These results are expected from the expansions of Eqs. \ref{maxans}
 and \ref{minansdef}, since
the terms which differ between the Improved Born Approximation and the
one-loop calculation are suppressed in the large $M_S$ limit.    

\begin{figure}[t]
\begin{center}
\includegraphics[scale=0.6]{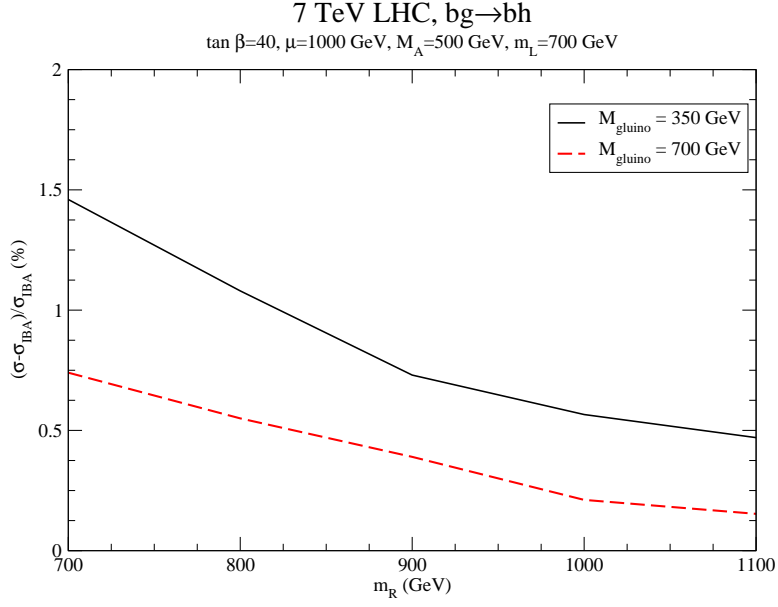} 
\caption[]{
Percentage difference between the Improved Born Approximation
and the exact one-loop SQCD calculation for $pp
\rightarrow bh$ as a function of $m_R$
 at $\sqrt{s}=7~ TeV$ and $\tan\beta=40$.
}
\label{fg:compsig2}
\end{center}
\end{figure}

Fig. \ref{fg:compsig1} compares the total SQCD rate for maximal and minimal
mixing, which bracket the allowed mixing possibilities.  For large $M_S$,
the effect of the mixing is quite small, while for $M_S\sim 800~GeV$, the 
mixing effects are at most a few $fb$.  The accuracy of the Improved Born
Approximation as a function of $m_R$ is shown in Fig. \ref{fg:compsig2}
for fixed $M_A,\mu$, and $m_L$. As $m_R$ is increased, the effects 
become very tiny.  Even for light gluino masses, the Improved
Born Approximation reproduces the exact SQCD result to within a few percent.

\begin{figure}[t]
\begin{center}
\includegraphics[scale=0.6]{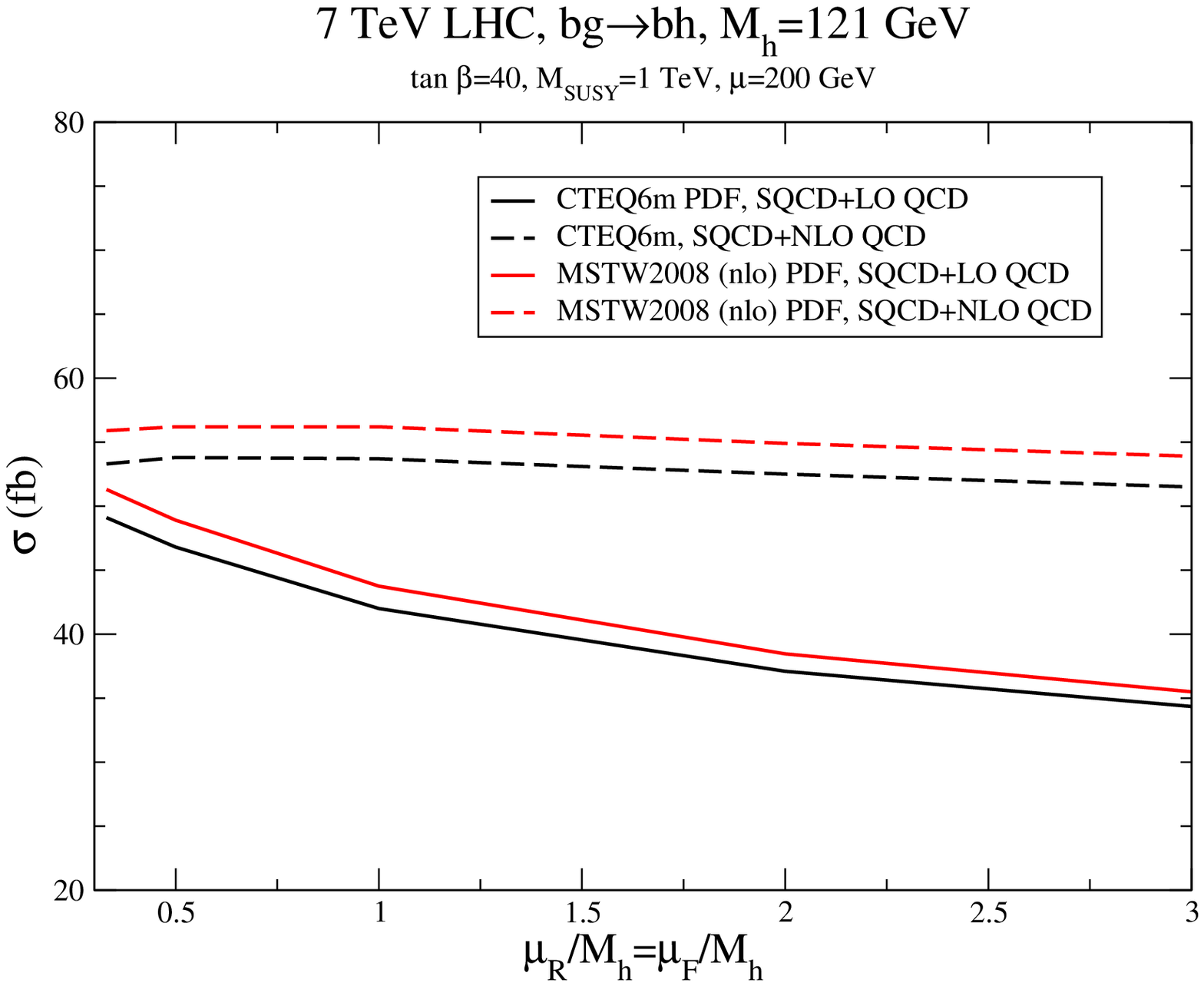} 
\caption[]{Total cross section for $pp\rightarrow b (\overline{b})h$
production including NLO QCD and SQCD corrections (dotted lines) as 
a function of renormalization/factorization scale using CTEQ6m (black) 
and MSTW2008 NLO (red) PDFs. We take $M_{\tilde g}=1~TeV$ and the
remaining MSSM parameters as in Fig. \ref{fg:maxmix}.}
\label{fg:susybh}
\end{center}
\end{figure}

\begin{figure}[t]
\begin{center}
\includegraphics[scale=0.6]{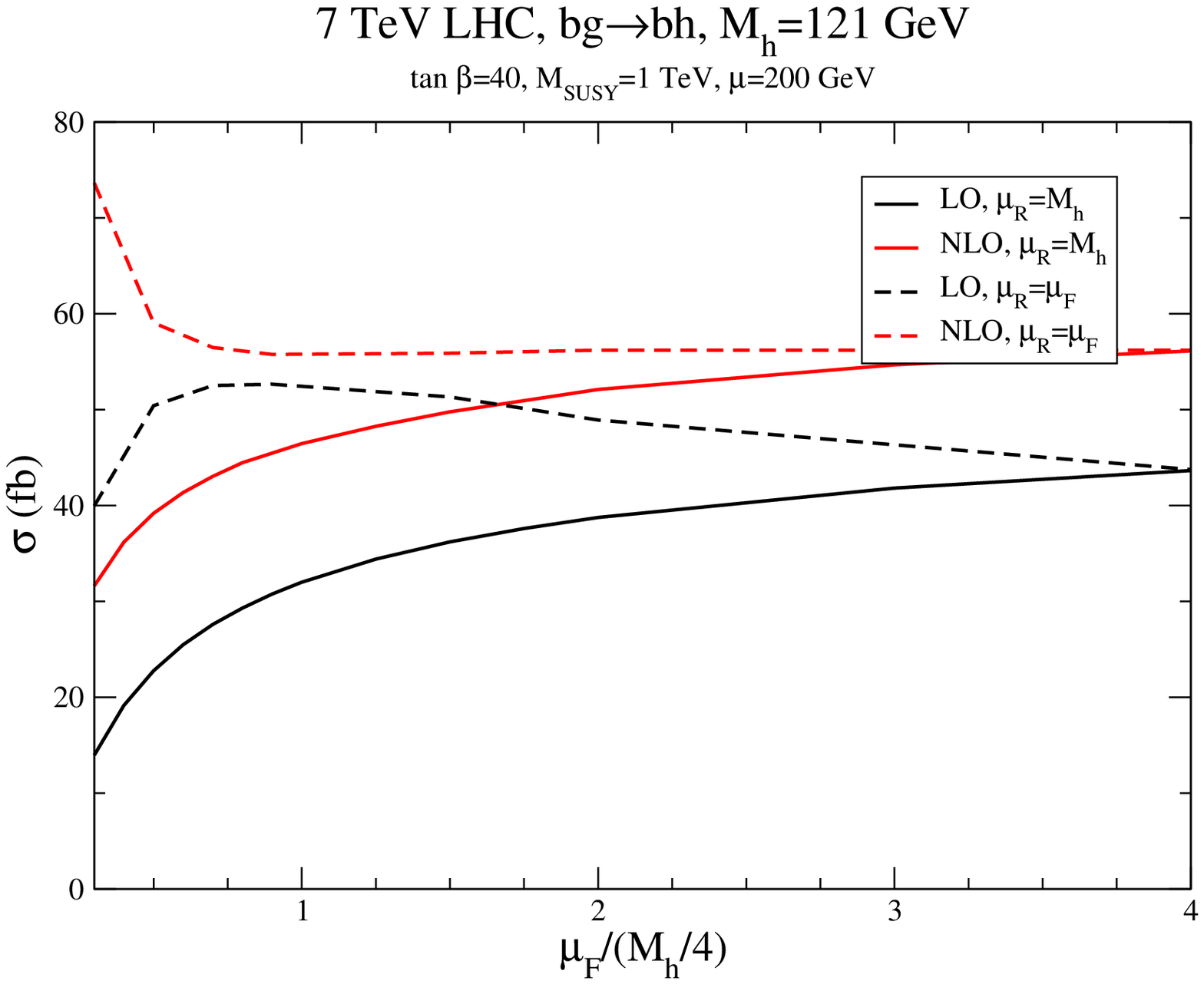} 
\caption[]{Total cross section for $pp\rightarrow b (\overline{b})h$
production including NLO QCD and SQCD corrections  as 
a function of the factorization scale using 
MSTW2008 NLO  PDFs. We take $M_{\tilde g}=1~TeV$ and the
remaining MSSM parameters as in Fig. \ref{fg:maxmix}.}
\label{fg:susybh_muf}
\end{center}
\end{figure}

In Fig. \ref{fg:susybh}, we show the scale dependence for
the total rate, including NLO QCD and SQCD corrections (dotted lines) for
a representative set of MSSM parameters at $\sqrt{s}=7~TeV$.  The
NLO scale dependence is quite small when $\mu_R=\mu_F\sim M_h$. 
 However, there is a roughly
$\sim 5\%$ difference between the predictions found 
using the CTEQ6m PDFs and
the MSTW2008 NLO PDFs\cite{Martin:2009iq}. In Fig. \ref{fg:susybh_muf},
we show the scale dependence for small $\mu_F$ (as preferred
 by \cite{Maltoni:2005wd}), and see that it is
significantly larger than in Fig. \ref{fg:susybh}. This is consistent with
the results of \cite{Harlander:2003ai,Dittmaier:2011ti}.

\section{Conclusion}
Our major results are the analytic expressions for the SQCD corrections to
$b$ Higgs associated production in the minimal (Eqs.~ \ref{dg2max},
\ref{dkapmax} and \ref{maxans}) and 
maximal (Eqs.~\ref{minkapdef}, \ref{dg2min} and \ref{minansdef}) $b$ squark
mixing scenarios for large $\tan\beta$ and squark masses,$M_S$.  
These results clearly 
demonstrate that deviations from the $\Delta_b$ approximation are
suppressed by powers of $(M_{EW}/M_S)$ in the large $\tan \beta$ region.
The $\Delta_b$ approximation hence yields an accurate prediction 
in the $5$ flavor number scheme for
the cross section for squark and gluino masses at the $TeV$ scale. 
As a by-product of our calculation, we update the predictions for $b$ Higgs
production at $\sqrt{s}=7~TeV$.

\noindent
\section*{Acknowledgements}
S.~Dawson~ and P.Jaiswal are supported by the United States Department
of Energy under Grant DE-AC02-98CH10886.

\noindent
\section*{Appendix A: Passarino-Veltman Functions}
\label{app:P-V}

The scalar integrals are defined as:
\begin{eqnarray}
\label{eq:A0}
{i\over 16\pi^2}A_0(M_0^2) &=& \int \frac{d^nk}{(2\pi)^n} \frac{1}{N_0}\,, 
\nonumber\\
\label{eq:B0}
{i\over 16\pi^2}B_0(p_1^2;M_0^2,M_1^2) &=& \int \frac{d^nk}{(2\pi)^n} \frac{1}{N_0 N_1}\,,
\nonumber\\
\label{eq:C0}
{i\over 16\pi^2}C_0(p_1^2,p_2^2,(p_1+p_2)^2;M_0^2,
M_1^2,M_2^2) &=& \int \frac{d^nk}{(2\pi)^n} \frac{1}{N_0 N_1 N_2}\,, 
\nonumber\\
\label{eq:D0}
{i\over 16\pi^2}D_0(p_1^2,p_2^2,p_3^2,p_4^2,(p_1+p_2)^2,
(p_2+p_3)^2;M_0^2,M_1^2,M_2^2,M_3^2) 
&&\nonumber \\
\qquad\qquad\qquad =\int \frac{d^nk}{(2\pi)^n} \frac{1}{N_0 N_1 N_2 N_3}\,, 
&&
\end{eqnarray}
where,
\begin{eqnarray}
N_0 &=& k^2 - M_0^2 
\nonumber\\
N_1 &=& (k + p_1)^2 - M_1^2 
\nonumber\\
N_2 &=& (k + p_1 + p_2)^2 - M_2^2 
\nonumber\\
N_3 &=& (k + p_1 + p_2 + p_3)^2 - M_3^2 \,.
\end{eqnarray}

The tensor integrals encountered are expanded in terms of the 
external momenta $p_i$ and the metric tensor $g^{\mu\nu}$.  For the two-point
function we write:
\begin{eqnarray}
{i\over 16\pi^2}B^\mu(p_1^2;M_0^2,M_1^2) 
&=& \int \frac{d^nk}{(2\pi)^n} \frac{k^\mu}{N_0 N_1} 
\nonumber\\
&\equiv& {i\over 16\pi^2}p_1^\mu B_1(p_1^2,M_0^2,M_1^2)\,,
\end{eqnarray}
while for the three-point functions we have both rank-one and rank-two tensor 
integrals which we expand as:
\begin{eqnarray}
C^\mu(p_1^2,p_2^2,(p_1+p_2)^2;M_0^2,M_1^2,M_2^2) 
&=& p_1^\mu C_{11} + p_2^\mu C_{12} \,,
\nonumber\\
C^{\mu\nu}(p_1^2,p_2^2,(p_1+p_2)^2;
M_0^2,M_1^2,M_2^2) &=& p_1^\mu p_1^\nu C_{21} + 
  p_2^\mu p_2^\nu C_{22} \nonumber\\
&+& (p_1^\mu p_2^\nu + p_1^\nu p_2^\mu) C_{23} + g^{\mu\nu} C_{24} \,,
\end{eqnarray} 
where:
\begin{equation}
{i\over 16\pi^2}
C^\mu (C^{\mu\nu})(p_1^2,p_2^2,(p_1+p_2)^2;
M_0^2,M_1^2,M_2^2)  \equiv
\int \frac{d^nk}{(2\pi)^n} \frac{k^\mu (k^\mu k^\nu)}{N_0 N_1 N_2}
\end{equation}

Finally, for the box diagrams, we encounter  rank-one and rank-two
 tensor integrals which
are written in terms of the Passarino-Veltmann coefficients as:
\begin{eqnarray}
{i\over 16\pi^2}D^\mu(p_1^2,p_2^2,p_3^2,p_4^2,(p_1+p_2)^2,(p_2+p3)^2
;M_0^2,M_1^2,M_2^2)  &\equiv&
\int \frac{d^nk}{(2\pi)^n} \frac{k^\mu}{N_0 N_1 N_2 N_3} 
\nonumber\\
\qquad\qquad\qquad ={i\over 16\pi^2}
\biggl\{ p_1^\mu D_{11} + p_2^\mu D_{12} + p_3^\mu D_{13}\biggr\} \,.
&&
\end{eqnarray}

\begin{eqnarray}
{i\over 16\pi^2}D^{\mu\nu}(p_1^2,p_2^2,p_3^2,
p_4^2,(p_1+p_2)^2,(p_2+p_3)^2;M_0^2,M_1^2,M_2^2)  &\equiv&
\int \frac{d^nk}{(2\pi)^n} 
\frac{k^\mu k^\nu}{N_0 N_1 N_2 N_3} 
\nonumber\\
\quad\qquad\qquad
= {i\over 16\pi^2}
\biggl\{ g^{\mu\nu} D_{00}
+{\hbox{tensor structures not needed here}}\biggr\} \,.
&&
\end{eqnarray}

\noindent
\section*{Appendix B: One-Loop Results}
In this appendix we give the
non-zero contributions of the individual diagrams in terms of
the basis functions of Eq. \ref{eq: SME}
and the decompositions of Eq. \ref{onedef}.
 The contributions proportional to $m_b \tan
\beta$  are new and 
were  not included in the results of Ref.\cite{Dawson:2007ur}.
Although we specialize to the case of the lightest Higgs boson, $h$,
our results are easily generalized to the heavier neutral Higgs boson, $H$,
and so the Feynman diagrams in this appendix are shown for $\phi_i=h,H$.

The self-energy diagrams of Fig. \ref{fg:self}:
\begin{figure}[hbtp!]
\begin{center}
\includegraphics[scale=0.7]
{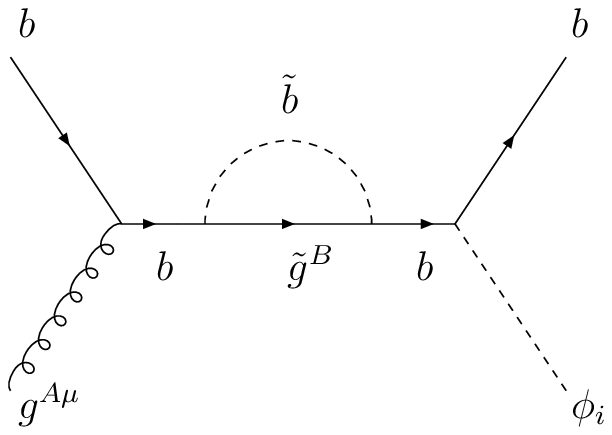}
\includegraphics[scale=0.7]
{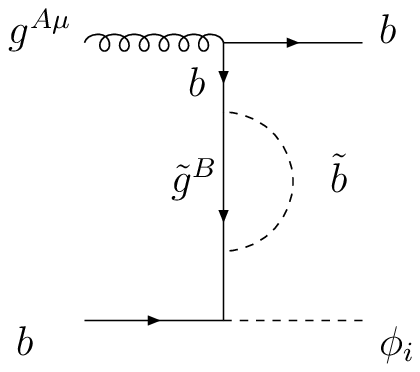}  
\vspace*{8pt}
\caption[]{Self-energy diagrams, $S_1$ and $S_2$.}
\label{fg:self}
\end{center}
\end{figure}

\begin{eqnarray}
X_{S_{1}}^{\left(t\right)} & = & \frac{4}{3}\sum_{i=1}^{2}\left\{ B_{1}-\left(-1\right)^{i}\frac{2m_{b}M_{\tilde{g}}s_{2\tilde{b}}}{t}B_{0}\right\} \left(M_{\tilde{b}_{i}}^{2}\right)\nonumber \\
X_{S_{1}}^{\left(2\right)} & = & -\frac{4}{3}\sum_{i=1}^{2}\left(-1\right)^{i}\frac{m_{b}M_{\tilde{g}}s_{2\tilde{b}}}{t}B_{0}\left(M_{\tilde{b}_{i}}^{2}\right)\label{eq: S1}
\end{eqnarray}
where we have have used the shorthand notation for the arguments of Passarino-Veltman
 functions, $
B_{0,1}\left(M_{\tilde{b}_{i}}^{2}\right)\equiv
B_{0,1}\left(t;M_{\tilde{g}}^{2},M_{\tilde{b}_{i}}^{2}\right)$.

\begin{eqnarray}
X_{S_{2}}^{\left(s\right)} & = & \frac{4}{3}\sum_{i=1}^{2}\left\{ B_{1}-\left(-1\right)^{i}\frac{2m_{b}M_{\tilde{g}}s_{2\tilde{b}}}{s}B_{0}\right\} \left(M_{\tilde{b}_{i}}^{2}\right)\nonumber \\
X_{S_{2}}^{\left(2\right)} & = & -\frac{4}{3}\sum_{i=1}^{2}\left(-1\right)^{i}\frac{m_{b}M_{\tilde{g}}s_{2\tilde{b}}}{s}B_{0}\left(M_{\tilde{b}_{i}}^{2}\right)\label{eq:S2}
\end{eqnarray}
and
$B_{0,1}\left(M_{\tilde{b}_{i}}^{2}\right)\equiv
B_{0,1}\left(s;M_{\tilde{g}}^{2},M_{\tilde{b}_{i}}^{2}\right)$

The vertex functions of Fig. \ref{fg:v12}:
\begin{figure}[hbtp!]
\begin{center}
\includegraphics[scale=0.7]
{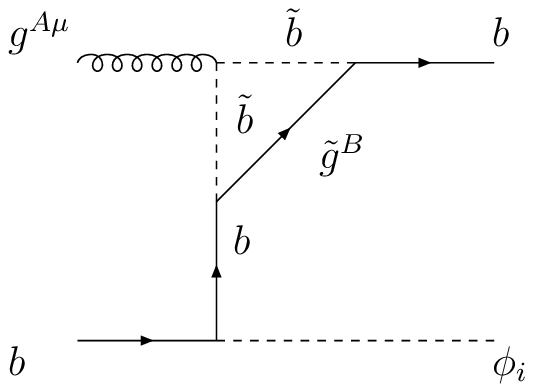}  
\includegraphics[scale=0.7]
{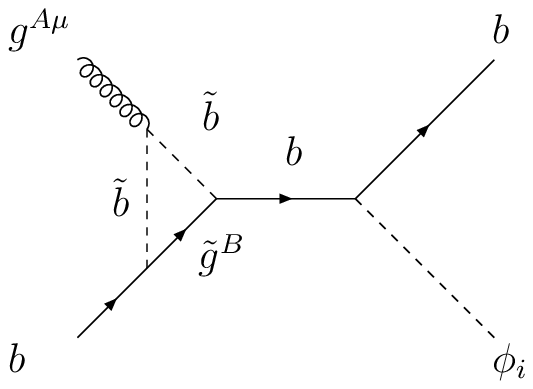}  
\vspace*{8pt}
\caption[]{Virtual diagrams, $V_1$ and $V_2$.}
\label{fg:v12}
\end{center}
\end{figure}

Diagram $V_1$:
\begin{eqnarray}
X_{V_{1}}^{\left(s\right)} & = & \frac{s}{6}\sum_{i=1}^{2}\left\{ C_{12}+C_{23}-\left(-1\right)^{i}\frac{2m_{b}M_{\tilde{g}}s_{2\tilde{b}}}{t}\left(C_{0}+C_{11}\right)\right\} \left(M_{\tilde{b_{i}}}^{2}\right)\nonumber \\
X_{V_{1}}^{\left(t\right)} & = & -\frac{1}{6}\sum_{i=1}^{2}\left\{ t\left(C_{12}+C_{23}\right)+2C_{24}-\left(-1\right)^{i}2m_{b}M_{\tilde{g}}s_{2\tilde{b}}\left(C_{0}+C_{11}\right)\right\} \left(M_{\tilde{b_{i}}}^{2}\right)\nonumber \\
X_{V_{1}}^{\left(1\right)} & = & -\frac{u}{3}\sum_{i=1}^{2}\left\{ C_{12}+C_{23}-\left(-1\right)^{i}\frac{2m_{b}M_{\tilde{g}}s_{2\tilde{b}}}{t}\left(C_{0}+C_{11}\right)\right\} \left(M_{\tilde{b_{i}}}^{2}\right)\nonumber \\
X_{V_{1}}^{\left(3\right)} & = & -\frac{1}{3}\sum_{i}\left(-1\right)^{i}m_{b}M_{\tilde{g}}s_{2\tilde{b}}\left(C_{0}+C_{11}\right)\left(M_{\tilde{b_{i}}}^{2}\right)\label{eq: V1}
\end{eqnarray}
where 
$
C_{0,11,12,23,24}\left(M_{\tilde{b}_{i}}^{2}\right)\equiv
C_{0,11,12,23,24}\left(0,0,t;M_{\tilde{g}}^{2},M_{\tilde{b_{i}}}^{2},M_{\tilde{b_{i}}}^{2}\right)$.

Diagram $V_{2}$:
\begin{eqnarray}
X_{V_{2}}^{\left(s\right)} & = & -\frac{1}{3}
\sum_{i=1}^{2}C_{24}\left(M_{\tilde{b}_i}^{2}\right)\nonumber \\
X_{V_{2}}^{\left(1\right)} & = & -\frac{u}{3}\sum_{i=1}^{2}\left\{ C_{12}+C_{23}-\left(-1\right)^{i}\frac{2m_{b}M_{\tilde{g}}s_{2\tilde{b}}}{s}\left(C_{0}+C_{11}\right)\right\} 
\left(M_{\tilde{b}_i}^{2}\right)\nonumber \\
X_{V_{2}}^{\left(4\right)} & = & \frac{1}{3}\sum_{i}\left(-1\right)^{i}m_{b}M_{\tilde{g}}s_{2\tilde{b}}\left(C_{0}+C_{11}\right)
\left(M_{\tilde{b}_i}^{2}\right)\label{eq: V2}
\end{eqnarray}
where 
$
C_{0,11,12,23,24}\left(M_{\tilde{b}_{i}}^{2}\right)\equiv
C_{0,11,12,23,24}\left(0,0,s;M_{\tilde{g}}^{2},M_{\tilde{b_{i}}}^{2},M_{\tilde{b_{i}}}^{2}\right)$.

The vertex functions of Fig. \ref{fg:v34}:
\begin{figure}[hbtp!]
\begin{center}
\includegraphics[scale=0.7]
{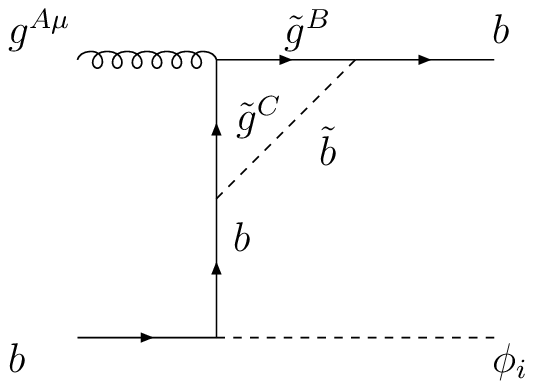}  
\includegraphics[scale=0.7]
{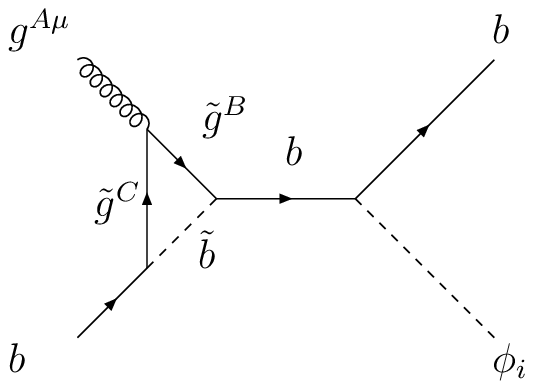}  
\vspace*{8pt}
\caption[]{Virtual diagrams, $V_3$ and $V_4$.}
\label{fg:v34}
\end{center}
\end{figure}

Diagram $V_{3}$:
\begin{eqnarray}
X_{V_{3}}^{\left(s\right)} & = & 
\frac{3s}{2}\sum_{i=1}^{2}\left\{ C_{12}
+C_{23}-\left(-1\right)^{i}
\frac{2m_{b}M_{\tilde{g}}s_{2\tilde{b}}}{t}
\left(C_{0}+C_{12}\right)\right\} 
\left(M_{\tilde{b_{i}}}^{2}\right)\nonumber \\
X_{V_{3}}^{\left(t\right)} & = & -\frac{3}{2}\sum_{i=1}^{2}\left\{ M_{\tilde{g}}^{2}C_{0}-2\left(1-\epsilon\right)C_{24}-\left(-1\right)^{i}2m_{b}M_{\tilde{g}}s_{2\tilde{b}}C_{12}\right\} \left(M_{\tilde{b_{i}}}^{2}\right)\nonumber \\
X_{V_{3}}^{\left(1\right)} & = & -3u\sum_{i=1}^{2}\left\{ C_{12}+C_{23}-\left(-1\right)^{i}\frac{2m_{b}M_{\tilde{g}}s_{2\tilde{b}}}{t}\left(C_{0}+C_{12}\right)\right\} \left(M_{\tilde{b_{i}}}^{2}\right)\nonumber \\
X_{V_{3}}^{\left(2\right)} & = & -\frac{3}{2}\sum_{i=1}^{2}\left(-1\right)^{i}m_{b}M_{\tilde{g}}s_{2\tilde{b}}C_{0}\left(M_{\tilde{b_{i}}}^{2}\right)\nonumber \\
X_{V_{3}}^{\left(3\right)} & = & -3\sum_{i=1}^{2}\left(-1\right)^{i}m_{b}M_{\tilde{g}}s_{2\tilde{b}}\left\{ C_{0}+C_{12}\right\} \left(M_{\tilde{b_{i}}}^{2}\right)\label{eq: V3}
\end{eqnarray}
where 
$
C_{0,11,12,23,24}\left(M_{\tilde{b}_{i}}^{2}\right)\equiv
C_{0,11,12,23,24}\left(0,0,t;M_{\tilde{g}}^{2},M_{\tilde{g}}^{2},M_{\tilde{b_{i}}}^{2}\right)$.

Diagram $V_{4}$:
\begin{eqnarray}
X_{V_{4}}^{\left(s\right)} & = & -\frac{3}{2}\sum_{i=1}^{2}\left\{ M_{\tilde{g}}^{2}C_{0}-2\left(1-\epsilon\right)C_{24}-s\left(C_{12}+C_{23}\right)+\left(-1\right)^{i}2m_{b}M_{\tilde{g}}s_{2\tilde{b}}C_{0}\right\} \left(M_{\tilde{b_{i}}}^{2}\right)\nonumber \\
X_{V_{4}}^{\left(1\right)} & = & -3u\sum_{i=1}^{2}\left\{ C_{12}+C_{23}-\left(-1\right)^{i}\frac{2m_{b}M_{\tilde{g}}s_{2\tilde{b}}}{s}\left(C_{0}+C_{12}\right)\right\} \left(M_{\tilde{b_{i}}}^{2}\right)\nonumber \\
X_{V_{4}}^{\left(2\right)} & = & -\frac{3}{2}\sum_{i=1}^{2}\left(-1\right)^{i}m_{b}M_{\tilde{g}}s_{2\tilde{b}}C_{0}\left(M_{\tilde{b_{i}}}^{2}\right)\nonumber \\
X_{V_{4}}^{\left(4\right)} & = & 3\sum_{i=1}^{2}\left(-1\right)^{i}m_{b}M_{\tilde{g}}s_{2\tilde{b}}\left\{ C_{0}+C_{12}\right\} \left(M_{\tilde{b_{i}}}^{2}\right)\label{eq: V4}
\end{eqnarray}
where 
$
C_{0,11,12,23,24}\left(M_{\tilde{b}_{i}}^{2}\right)\equiv
C_{0,11,12,23,24}\left(0,0,s;M_{\tilde{g}}^{2},M_{\tilde{g}}^{2},M_{\tilde{b_{i}}}^{2}\right)$.

The vertex functions of Fig. \ref{fg:v56}:
\begin{figure}[hbtp!]
\begin{center}
\includegraphics[scale=0.7]
{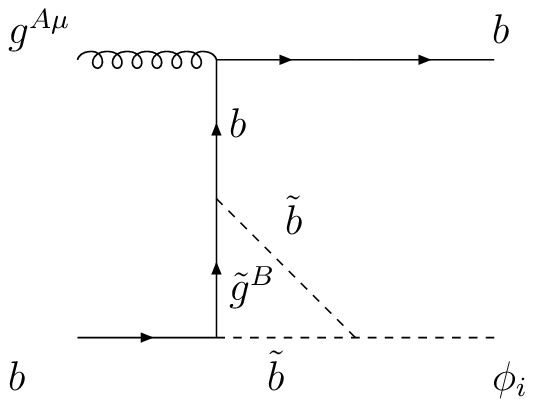}  
\includegraphics[scale=0.7]
{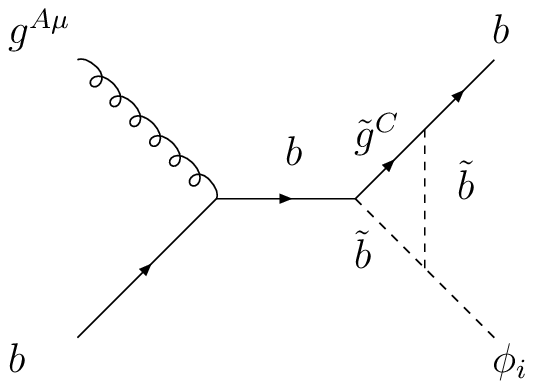}  
\vspace*{8pt}
\caption[]{Virtual diagrams, $V_5$ and $V_6$.}
\label{fg:v56}
\end{center}
\end{figure}

Diagram 
$V_{5}$:
\begin{eqnarray}
X_{V_{5}}^{\left(t\right)} & = & \frac{4}{3}\sum_{i,j=1}^{2}C_{h,ij}\left\{ \delta_{ij}m_{b}C_{11}+a_{ij}M_{\tilde{g}}C_{0}\right\} \left(M_{\tilde{b}_{i}}^{2},M_{\tilde{b}_{j}}^{2}\right)\nonumber \\
X_{V_{5}}^{\left(2\right)} & = & \frac{4}{3}m_{b}\sum_{i,j=1,2}C_{h,ij}\delta_{ij}C_{12}\left(M_{\tilde{b}_{i}}^{2},M_{\tilde{b}_{j}}^{2}\right)\label{eq: V5}
\end{eqnarray}
where 
$
C_{0,11,12,23,24}\left(M_{\tilde{b}_{i}}^{2},M_{\tilde{b}_{j}}^{2}\right)\equiv
C_{0,11,12,23,24}\left(0,
M_{h}^{2},t;M_{\tilde{g}}^{2},M_{\tilde{b}_{i}}^{2},M_{\tilde{b}_{j}}^{2}\right)\, ,
$
the squark mixing matrix is defined, 
\begin{equation}
\left(\begin{array}{cc}
a_{11} & a_{12}\\
a_{21} & a_{22}\end{array}
\right)=
\left(
\begin{array}{cc}
 s_{2\tilde{b}}& c_{2\tilde{b}}\\
c_{2\tilde{b}} & 
-s_{2\tilde{b}}\end{array}\right)\label{eq: a}\end{equation}
and 
the light Higgs-squark-squark couplings $C_{h,ij}$,
 are normalized with respect to the Higgs-quark-quark 
coupling\cite{Gunion:1989we},
\begin{eqnarray}
C_{h,11}+C_{h,22} & = & 4m_{b}+\frac{2M_{Z}^{2}}{m_{b}}I_{3}^{b}
\frac{s_{\alpha+\beta}c_{\beta}}{s_{\alpha}}\\
C_{h,11}-C_{h,22} & = & 2Y_{b}s_{2\tilde{b}}
+\frac{2M_{Z}^{2}}{m_{b}}c_{2\tilde{b}}\left(I_{3}^{b}-2Q_{b}s_{W}^{2}
\right)\frac{s_{\alpha+\beta}c_{\beta}}{s_{\alpha}}\\
C_{h,12}=C_{h,21} & = & Y_{b}c_{2\tilde{b}}-\frac{M_{Z}^{2}}
{m_{b}}s_{2\tilde{b}}\left(I_{3}^{b}-2Q^{b}s_{W}^{2}\right)
\frac{s_{\alpha+\beta}c_{\beta}}{s_{\alpha}}
\, ,
\end{eqnarray}
$s_W^2=\sin \theta_W^2=1-M_W^2/M_Z^2$ and $Y_b$ is defined below
Eq. \ref{dg2max}.

Diagram $V_{6}$:
\begin{eqnarray}
X_{V_{6}}^{\left(s\right)} & = & \frac{4}{3}\sum_{i,j=1,2}C_{h,ij}\left\{ \delta_{ij}m_{b}C_{11}+a_{ij}M_{\tilde{g}}C_{0}\right\} \left(M_{\tilde{b}_{i}}^{2},M_{\tilde{b}_{j}}^{2}\right)\nonumber \\
X_{V_{6}}^{\left(2\right)} & = & \frac{4}{3}m_{b}\sum_{i,j=1,2}C_{h,ij}\delta_{ij}C_{12}\left(M_{\tilde{b}_{i}}^{2},M_{\tilde{b}_{j}}^{2}\right)\nonumber \\
X_{V_{6}}^{\left(t\right)} & = & X_{V_{6}}^{\left(3\right)}=X_{V_{6}}^{\left(4\right)}=0\label{eq: V6}
\end{eqnarray}
where 
$
C_{0,11,12,23,24}\left(M_{\tilde{b}_{i}}^{2},M_{\tilde{b}_{j}}^{2}\right)\equiv
C_{0,11,12,23,24}\left(0,M_{h}^{2},s;M_{\tilde{g}}^{2},M_{\tilde{b}_{i}}^{2},M_{\tilde{b}_{j}}^{2}\right)
.$

The box diagram of Fig. \ref{fg:box1}:
\begin{figure}[hbtp!]
\begin{center}
\includegraphics[scale=0.7]
{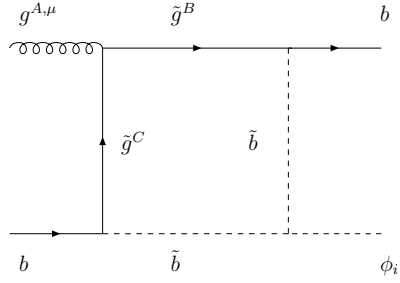}  
\vspace*{8pt}
\caption[]{Box diagram, $B_1$.}
\label{fg:box1}
\end{center}
\end{figure}

\begin{eqnarray}
X_{B_{1}}^{\left(s\right)} & = & \frac{3M_{\tilde{g}}s}{2}
\sum_{i,j=1,2}a_{ij}C_{h,ij}\left\{ D_{0}+D_{13}\right\} \left(M_{\tilde{b}_{i}}^{2},M_{\tilde{b}_{j}}^{2}\right)\nonumber \\
X_{B_{1}}^{\left(t\right)} & = &
 -\frac{3M_{\tilde{g}}t}{2}\sum_{i,j=1,2}a_{ij}C_{h,ij}
D_{13}\left(M_{\tilde{b}_{i}}^{2},M_{\tilde{b}_{j}}^{2}\right)\nonumber \\
X_{B_{1}}^{\left(1\right)} & 
= & 3M_{\tilde{g}}u\sum_{i,j=1,2}a_{ij}C_{h,ij}
\left\{ D_{11}-D_{13}\right\} \left(M_{\tilde{b}_{i}}^{2},M_{\tilde{b}_{j}}^{2}\right)\nonumber \\
X_{B_{1}}^{\left(2\right)} & = & -\frac{3m_{b}}{2}\sum_{i,j=1,2}\delta_{ij}C_{h,ij}\left\{ M_{\tilde{g}}^{2}D_{0}-2D_{00}\right\} \left(M_{\tilde{b}_{i}}^{2},M_{\tilde{b}_{j}}^{2}\right)\label{eq:B1}
\end{eqnarray}
where,
$
D_0\left(M_{\tilde{b}_{i}}^{2},M_{\tilde{b}_{j}}^{2}\right)
\equiv
D_0\left(0,0,0,M_h^2,s,t;M_{\tilde{b}_{i}}^{2},M_{\tilde{g}}^{2},M_{\tilde{g}}^{2},M_{\tilde{b}_{j}}^{2}\right)$.

The box diagram of Fig. \ref{fg:box2}:
\begin{figure}[hbtp!]
\begin{center}
\includegraphics[scale=0.7]
{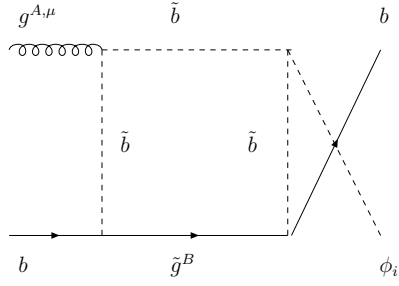}  
\vspace*{8pt}
\caption[]{Box diagram, $B_2$.}
\label{fg:box2}
\end{center}
\end{figure}

Diagram $B_{2}$:
\begin{eqnarray}
X_{B_{2}}^{\left(s\right)} & = & 
-\frac{M_{\tilde{g}}s}{6}\sum_{i,j=1,2}a_{ij}C_{h,ij}
\left\{ D_{0}+D_{11}\right\} \left(M_{\tilde{b}_{i}}^{2},M_{\tilde{b}_{j}}^{2}\right)\nonumber \\
X_{B_{2}}^{\left(t\right)} & = & \frac{M_{\tilde{g}}t}{6}\sum_{i,j=1,2}a_{ij}C_{h,ij}
\left\{ D_{0}+D_{11}\right\} \left(M_{\tilde{b}_{i}}^{2},M_{\tilde{b}_{j}}^{2}\right)\nonumber \\
X_{B_{2}}^{\left(1\right)} & = & \frac{M_{\tilde{g}}u}{3}\sum_{i,j=1,2}a_{ij}C_{h,ij}
\left\{ D_{11}-D_{12}\right\} \left(M_{\tilde{b}_{i}}^{2},M_{\tilde{b}_{j}}^{2}\right)\nonumber \\
X_{B_{2}}^{\left(2\right)} & = & -\frac{m_{b}}{3}\sum_{i,j=1,2}\delta_{ij}C_{h,ij}D_{00}\left(M_{\tilde{b}_{i}}^{2},M_{\tilde{b}_{j}}^{2}\right)\label{eq:B2}
\end{eqnarray}
where
$
D_0\left(M_{\tilde{b}_{i}}^{2},M_{\tilde{b}_{j}}^{2}\right)\equiv
D_0\left(0,0,0,M_h^2,u,s;M_{\tilde{b}_{i}}^{2},M_{\tilde{g}}^{2},M_{\tilde{b}_{j}}^{2},M_{\tilde{b}_{j}}^{2}\right)$.

The box diagram of Fig. \ref{fg:box3}:
\begin{figure}[hbtp!]
\begin{center}
\includegraphics[scale=0.7]
{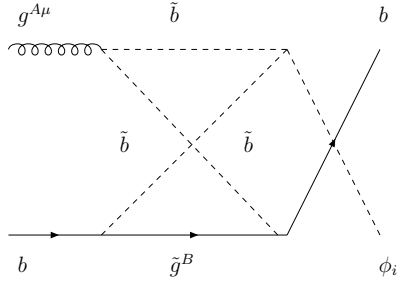}  
\vspace*{8pt}
\caption[]{Box diagram, $B_3$.}
\label{fg:box3}
\end{center}
\end{figure}

Diagram $B_{3}$:
\begin{eqnarray}
X_{B_{3}}^{\left(s\right)} & = & 
\frac{M_{\tilde{g}}s}{6}\sum_{i,j=1,2}a_{ij}C_{h,ij}
\left\{ D_{0}+D_{12}\right\} \left(M_{\tilde{b}_{i}}^{2},M_{\tilde{b}_{j}}^{2}\right)\nonumber \\
X_{B_{3}}^{\left(t\right)} & = & -\frac{M_{\tilde{g}}t}{6}\sum_{i,j=1,2}a_{ij}C_{h,ij}
\left\{ D_{0}+D_{12}\right\} \left(M_{\tilde{b}_{i}}^{2},M_{\tilde{b}_{j}}^{2}\right)\nonumber \\
X_{B_{3}}^{\left(1\right)} & 
= & \frac{M_{\tilde{g}}u}{3}\sum_{i,j=1,2}a_{ij}C_{h,ij}
\left\{ D_{11}-D_{12}\right\} \left(M_{\tilde{b}_{i}}^{2},M_{\tilde{b}_{j}}^{2}\right)\nonumber \\
X_{B_{3}}^{\left(2\right)} & = & -\frac{m_{b}}{3}\sum_{i,j=1,2}\delta_{ij}C_{h,ij}D_{00}\left(M_{\tilde{b}_{i}}^{2},M_{\tilde{b}_{j}}^{2}\right)
\label{eq:B3}
\end{eqnarray}
where
$
D_0\left(M_{\tilde{b}_{i}}^{2},M_{\tilde{b}_{j}}^{2}\right)\equiv
D_0\left(0,0,0,M_h^2,u,t;M_{\tilde{b}_{i}}^{2},M_{\tilde{g}}^{2},M_{\tilde{b}_{j}}^{2},M_{\tilde{b}_{j}}^{2}\right)$.

The vertex and external wavefunction 
counter terms, Eq. \ref{eq:delZ}, along with the subtraction of 
Eq. \ref{ctdef}, give the counterterm of Eq. \ref{cttot}: 
\begin{eqnarray}
X_{CT}^{\left(s\right)} & = &X_{CT}^{\left(t\right)}
=\biggl({4\pi\over \alpha_s(\mu_R)}\biggr)
\biggl[\delta Z_b^V+{\delta m_b\over m_b}+\delta_{CT}\biggr]
\nonumber \\&=& 
\frac{4}{3}\biggl[
2M_{\tilde g} Y_b I(M_{{\tilde b}_1}, M_{{\tilde b}_2}, M_{\tilde g})
+
\sum_{i=1}^{2}
\biggl(
-\left(-1\right)^{i}2 m_{b} s_{2\tilde{b}}B_{0}^\prime
+2m_b^2 B_1^\prime\biggr)(0;M_{\tilde g}^2, M_{{\tilde b}_i}^2)\biggr]
\, .
\label{eq:ct 1}
\end{eqnarray}
Note that the counterterm contains no large $\tan\beta$ enhanced contribution.

\noindent
\section*{Appendix C: Definitions}
In this appendix we define the f
unctions used in the expansions of the Passarino-Veltman integrals in 
the maximum and minimum mixing scenarios, where 
$R\equiv {M_{\tilde g}\over M_S}$ in the
maximal mixing scenario, and 
 $R_i\equiv{ M_{{\tilde b}_i}\over M_S}$ in the minimal mixing scenario:
\begin{eqnarray}
f_{1}\left(R\right) & = & \frac{2}{\left(1-R^{2}\right)^{2}}\left[1-R^{2}+R^{2}\log R^{2}\right]
\nonumber \\
f_{2}\left(R\right) & = & \frac{3}{\left(1-R^{2}\right)^{3}}\left[1-R^{4}+2R^{2}\log R^{2}\right]
\nonumber \\
f_{3}\left(R\right) & = & \frac{4}{\left(1-R^{2}\right)^{4}}\left[1+\frac{3}{2}R^{2}-3R^{4}+\frac{1}{2}R^{6}+3R^{2}\log R^{2}\right]
\nonumber \\
f_{4}\left(R\right) & = & \frac{5}{\left(1-R^{2}\right)^{5}}\left[\frac{1}{2}-4R^{2}+4R^{6}-\frac{1}{2}R^{8}-6R^{4}\log R^{2}\right]
\nonumber \\
h_{1}\left(R_{1},R_{2},n\right) & = & \left(\frac{R_{1}^{2}}{1-R_{1}^{2}}\right)^{n}\frac{\log R_{1}^{2}}{1-R_{1}^{2}}-\left(\frac{R_{2}^{2}}{1-R_{2}^{2}}\right)^{n}\frac{\log R_{2}^{2}}{1-R_{2}^{2}}
\nonumber \\
 &  & -\sum_{j=0}^{n}(-1)^j\frac{j+2}{2}\left\{ \left(1-R_{1}^{2}\right)^{j-n}-\left(1-R_{2}^{2}\right)^{j-n}\right\} 
\nonumber \\
h_{2}\left(R_{1},R_{2}\right) & = & \frac{R_{1}^{2}+R_{2}^{2}-2}{\left(1-R_{1}^{2}\right)\left(1-R_{2}^{2}\right)}+\frac{1}{R_{1}^{2}-R_{2}^{2}}\Biggl[\frac{R_{1}^{2}+R_{2}^{2}-2R_{1}^{4}}{\left(1-R_{1}^{2}\right)^{2}}\log R_{1}^{2}\nonumber \\
 &  & -\frac{R_{1}^{2}+R_{2}^{2}-2R_{2}^{4}}{\left(1-R_{2}^{2}\right)^{2}}\log R_{2}^{2}\Biggr]
\, .
\end{eqnarray}

Further, 
\begin{eqnarray}
f_{i}'\left(R\right) & \equiv & \frac{\mathrm{d}f_{i}\left(x\right)}{\mathrm{d}x^{2}}\Biggr|_{x=R}\nonumber \\
f_{i}^{-1}\left(R\right) & \equiv & \frac{f_{i}\left(1/R\right)}{R^{2}}\nonumber \\
\hat{f}_{i}\left(R\right) & \equiv & \frac{1}{R^{4}}\frac{\mathrm{d}f_{i}\left(x\right)}{\mathrm{d}x^{2}}\Biggr|_{x=1/R}\label{eq:func 2}
\, .
\end{eqnarray}

\bibliography{mssm_decoup}

\begin{thebibliography}{47}
\expandafter\ifx\csname natexlab\endcsname\relax\def\natexlab#1{#1}\fi
\expandafter\ifx\csname bibnamefont\endcsname\relax
  \def\bibnamefont#1{#1}\fi
\expandafter\ifx\csname bibfnamefont\endcsname\relax
  \def\bibfnamefont#1{#1}\fi
\expandafter\ifx\csname citenamefont\endcsname\relax
  \def\citenamefont#1{#1}\fi
\expandafter\ifx\csname url\endcsname\relax
  \def\url#1{\texttt{#1}}\fi
\expandafter\ifx\csname urlprefix\endcsname\relax\def\urlprefix{URL }\fi
\providecommand{\bibinfo}[2]{#2}
\providecommand{\eprint}[2][]{\url{#2}}

\bibitem[{\citenamefont{Djouadi}(2005)}]{Djouadi:2005gj}
\bibinfo{author}{\bibfnamefont{A.}~\bibnamefont{Djouadi}}
  (\bibinfo{year}{2005}), \eprint{hep-ph/0503173}.

\bibitem[{\citenamefont{Gunion et~al.}(1990)\citenamefont{Gunion, Haber, Kane,
  and Dawson}}]{Gunion:1989we}
\bibinfo{author}{\bibfnamefont{J.~F.} \bibnamefont{Gunion}},
  \bibinfo{author}{\bibfnamefont{H.~E.} \bibnamefont{Haber}},
  \bibinfo{author}{\bibfnamefont{G.~L.} \bibnamefont{Kane}}, \bibnamefont{and}
  \bibinfo{author}{\bibfnamefont{S.}~\bibnamefont{Dawson}},
  \emph{\bibinfo{title}{THE HIGGS HUNTER'S GUIDE}} (\bibinfo{publisher}{Addison
  Wesley (Menlo Park)}, \bibinfo{year}{1990}).

\bibitem[{\citenamefont{Carena and Haber}(2003)}]{Carena:2002es}
\bibinfo{author}{\bibfnamefont{M.~S.} \bibnamefont{Carena}} \bibnamefont{and}
  \bibinfo{author}{\bibfnamefont{H.~E.} \bibnamefont{Haber}},
  \bibinfo{journal}{Prog. Part. Nucl. Phys.} \textbf{\bibinfo{volume}{50}},
  \bibinfo{pages}{63} (\bibinfo{year}{2003}), \eprint{hep-ph/0208209}.

\bibitem[{\citenamefont{Benjamin et~al.}(2010)}]{Benjamin:2010xb}
\bibinfo{author}{\bibfnamefont{D.}~\bibnamefont{Benjamin}} \bibnamefont{et~al.}
  (\bibinfo{collaboration}{Tevatron New Phenomena and Higgs Working Group})
  (\bibinfo{year}{2010}), \eprint{1003.3363}.

\bibitem[{\citenamefont{Chatrchyan et~al.}(2011)}]{Chatrchyan:2011nx}
\bibinfo{author}{\bibfnamefont{S.}~\bibnamefont{Chatrchyan}}
  \bibnamefont{et~al.} (\bibinfo{collaboration}{CMS}) (\bibinfo{year}{2011}),
  \eprint{1104.1619}.

\bibitem[{\citenamefont{Aad et~al.}(2009)}]{Aad:2009wy}
\bibinfo{author}{\bibfnamefont{G.}~\bibnamefont{Aad}} \bibnamefont{et~al.}
  (\bibinfo{collaboration}{The ATLAS}) (\bibinfo{year}{2009}),
  \eprint{0901.0512}.

\bibitem[{\citenamefont{Bayatian et~al.}(2007)}]{Ball:2007zza}
\bibinfo{author}{\bibfnamefont{G.~L.} \bibnamefont{Bayatian}}
  \bibnamefont{et~al.} (\bibinfo{collaboration}{CMS}), \bibinfo{journal}{J.
  Phys.} \textbf{\bibinfo{volume}{G34}}, \bibinfo{pages}{995}
  (\bibinfo{year}{2007}).

\bibitem[{\citenamefont{Dawson et~al.}(2005)\citenamefont{Dawson, Jackson,
  Reina, and Wackeroth}}]{Dawson:2004sh}
\bibinfo{author}{\bibfnamefont{S.}~\bibnamefont{Dawson}},
  \bibinfo{author}{\bibfnamefont{C.~B.} \bibnamefont{Jackson}},
  \bibinfo{author}{\bibfnamefont{L.}~\bibnamefont{Reina}}, \bibnamefont{and}
  \bibinfo{author}{\bibfnamefont{D.}~\bibnamefont{Wackeroth}},
  \bibinfo{journal}{Phys. Rev. Lett.} \textbf{\bibinfo{volume}{94}},
  \bibinfo{pages}{031802} (\bibinfo{year}{2005}), \eprint{hep-ph/0408077}.

\bibitem[{\citenamefont{Dawson et~al.}(2006)\citenamefont{Dawson, Jackson,
  Reina, and Wackeroth}}]{Dawson:2005vi}
\bibinfo{author}{\bibfnamefont{S.}~\bibnamefont{Dawson}},
  \bibinfo{author}{\bibfnamefont{C.~B.} \bibnamefont{Jackson}},
  \bibinfo{author}{\bibfnamefont{L.}~\bibnamefont{Reina}}, \bibnamefont{and}
  \bibinfo{author}{\bibfnamefont{D.}~\bibnamefont{Wackeroth}},
  \bibinfo{journal}{Mod. Phys. Lett.} \textbf{\bibinfo{volume}{A21}},
  \bibinfo{pages}{89} (\bibinfo{year}{2006}), \eprint{hep-ph/0508293}.

\bibitem[{\citenamefont{Carena et~al.}(2007)\citenamefont{Carena, Menon, and
  Wagner}}]{Carena:2007aq}
\bibinfo{author}{\bibfnamefont{M.~S.} \bibnamefont{Carena}},
  \bibinfo{author}{\bibfnamefont{A.}~\bibnamefont{Menon}}, \bibnamefont{and}
  \bibinfo{author}{\bibfnamefont{C.~E.~M.} \bibnamefont{Wagner}},
  \bibinfo{journal}{Phys. Rev.} \textbf{\bibinfo{volume}{D76}},
  \bibinfo{pages}{035004} (\bibinfo{year}{2007}), \eprint{arXiv:0704.1143
  [hep-ph]}.

\bibitem[{\citenamefont{Campbell et~al.}(2004)}]{Campbell:2004pu}
\bibinfo{author}{\bibfnamefont{J.}~\bibnamefont{Campbell}} \bibnamefont{et~al.}
  (\bibinfo{year}{2004}), \eprint{hep-ph/0405302}.

\bibitem[{\citenamefont{Dittmaier et~al.}(2004)\citenamefont{Dittmaier, Kramer,
  and Spira}}]{Dittmaier:2003ej}
\bibinfo{author}{\bibfnamefont{S.}~\bibnamefont{Dittmaier}},
  \bibinfo{author}{\bibfnamefont{M.}~\bibnamefont{Kramer}}, \bibnamefont{and}
  \bibinfo{author}{\bibfnamefont{M.}~\bibnamefont{Spira}},
  \bibinfo{journal}{Phys. Rev.} \textbf{\bibinfo{volume}{D70}},
  \bibinfo{pages}{074010} (\bibinfo{year}{2004}), \eprint{hep-ph/0309204}.

\bibitem[{\citenamefont{Carena et~al.}(1999)\citenamefont{Carena, Mrenna, and
  Wagner}}]{Carena:1998gk}
\bibinfo{author}{\bibfnamefont{M.~S.} \bibnamefont{Carena}},
  \bibinfo{author}{\bibfnamefont{S.}~\bibnamefont{Mrenna}}, \bibnamefont{and}
  \bibinfo{author}{\bibfnamefont{C.~E.~M.} \bibnamefont{Wagner}},
  \bibinfo{journal}{Phys. Rev.} \textbf{\bibinfo{volume}{D60}},
  \bibinfo{pages}{075010} (\bibinfo{year}{1999}), \eprint{hep-ph/9808312}.

\bibitem[{\citenamefont{Dawson et~al.}(2004)\citenamefont{Dawson, Jackson,
  Reina, and Wackeroth}}]{Dawson:2003kb}
\bibinfo{author}{\bibfnamefont{S.}~\bibnamefont{Dawson}},
  \bibinfo{author}{\bibfnamefont{C.~B.} \bibnamefont{Jackson}},
  \bibinfo{author}{\bibfnamefont{L.}~\bibnamefont{Reina}}, \bibnamefont{and}
  \bibinfo{author}{\bibfnamefont{D.}~\bibnamefont{Wackeroth}},
  \bibinfo{journal}{Phys. Rev.} \textbf{\bibinfo{volume}{D69}},
  \bibinfo{pages}{074027} (\bibinfo{year}{2004}), \eprint{hep-ph/0311067}.

\bibitem[{\citenamefont{Maltoni et~al.}(2003)\citenamefont{Maltoni, Sullivan,
  and Willenbrock}}]{Maltoni:2003pn}
\bibinfo{author}{\bibfnamefont{F.}~\bibnamefont{Maltoni}},
  \bibinfo{author}{\bibfnamefont{Z.}~\bibnamefont{Sullivan}}, \bibnamefont{and}
  \bibinfo{author}{\bibfnamefont{S.}~\bibnamefont{Willenbrock}},
  \bibinfo{journal}{Phys. Rev.} \textbf{\bibinfo{volume}{D67}},
  \bibinfo{pages}{093005} (\bibinfo{year}{2003}), \eprint{hep-ph/0301033}.

\bibitem[{\citenamefont{Maltoni et~al.}(2005)\citenamefont{Maltoni, McElmurry,
  and Willenbrock}}]{Maltoni:2005wd}
\bibinfo{author}{\bibfnamefont{F.}~\bibnamefont{Maltoni}},
  \bibinfo{author}{\bibfnamefont{T.}~\bibnamefont{McElmurry}},
  \bibnamefont{and}
  \bibinfo{author}{\bibfnamefont{S.}~\bibnamefont{Willenbrock}},
  \bibinfo{journal}{Phys. Rev.} \textbf{\bibinfo{volume}{D72}},
  \bibinfo{pages}{074024} (\bibinfo{year}{2005}), \eprint{hep-ph/0505014}.

\bibitem[{\citenamefont{Dicus et~al.}(1999)\citenamefont{Dicus, Stelzer,
  Sullivan, and Willenbrock}}]{Dicus:1998hs}
\bibinfo{author}{\bibfnamefont{D.}~\bibnamefont{Dicus}},
  \bibinfo{author}{\bibfnamefont{T.}~\bibnamefont{Stelzer}},
  \bibinfo{author}{\bibfnamefont{Z.}~\bibnamefont{Sullivan}}, \bibnamefont{and}
  \bibinfo{author}{\bibfnamefont{S.}~\bibnamefont{Willenbrock}},
  \bibinfo{journal}{Phys. Rev.} \textbf{\bibinfo{volume}{D59}},
  \bibinfo{pages}{094016} (\bibinfo{year}{1999}), \eprint{hep-ph/9811492}.

\bibitem[{\citenamefont{Campbell et~al.}(2003)\citenamefont{Campbell, Ellis,
  Maltoni, and Willenbrock}}]{Campbell:2002zm}
\bibinfo{author}{\bibfnamefont{J.}~\bibnamefont{Campbell}},
  \bibinfo{author}{\bibfnamefont{R.~K.} \bibnamefont{Ellis}},
  \bibinfo{author}{\bibfnamefont{F.}~\bibnamefont{Maltoni}}, \bibnamefont{and}
  \bibinfo{author}{\bibfnamefont{S.}~\bibnamefont{Willenbrock}},
  \bibinfo{journal}{Phys. Rev.} \textbf{\bibinfo{volume}{D67}},
  \bibinfo{pages}{095002} (\bibinfo{year}{2003}), \eprint{hep-ph/0204093}.

\bibitem[{\citenamefont{Dittmaier et~al.}(2011)}]{Dittmaier:2011ti}
\bibinfo{author}{\bibfnamefont{S.}~\bibnamefont{Dittmaier}}
  \bibnamefont{et~al.} (\bibinfo{collaboration}{LHC Higgs Cross Section Working
  Group}) (\bibinfo{year}{2011}), \eprint{1101.0593}.

\bibitem[{\citenamefont{Field et~al.}(2007)\citenamefont{Field, Reina, and
  Jackson}}]{Field:2007ye}
\bibinfo{author}{\bibfnamefont{B.}~\bibnamefont{Field}},
  \bibinfo{author}{\bibfnamefont{L.}~\bibnamefont{Reina}}, \bibnamefont{and}
  \bibinfo{author}{\bibfnamefont{C.~B.} \bibnamefont{Jackson}},
  \bibinfo{journal}{Phys. Rev.} \textbf{\bibinfo{volume}{D76}},
  \bibinfo{pages}{074008} (\bibinfo{year}{2007}), \eprint{0705.0035}.

\bibitem[{\citenamefont{Dawson and Jaiswal}(2010)}]{Dawson:2010yz}
\bibinfo{author}{\bibfnamefont{S.}~\bibnamefont{Dawson}} \bibnamefont{and}
  \bibinfo{author}{\bibfnamefont{P.}~\bibnamefont{Jaiswal}},
  \bibinfo{journal}{Phys. Rev.} \textbf{\bibinfo{volume}{D81}},
  \bibinfo{pages}{073008} (\bibinfo{year}{2010}), \eprint{1002.2672}.

\bibitem[{\citenamefont{Beccaria et~al.}(2010)}]{Beccaria:2010fg}
\bibinfo{author}{\bibfnamefont{M.}~\bibnamefont{Beccaria}}
  \bibnamefont{et~al.}, \bibinfo{journal}{Phys. Rev.}
  \textbf{\bibinfo{volume}{D82}}, \bibinfo{pages}{093018}
  (\bibinfo{year}{2010}), \eprint{1005.0759}.

\bibitem[{\citenamefont{Dawson and Jackson}(2008)}]{Dawson:2007ur}
\bibinfo{author}{\bibfnamefont{S.}~\bibnamefont{Dawson}} \bibnamefont{and}
  \bibinfo{author}{\bibfnamefont{C.~B.} \bibnamefont{Jackson}},
  \bibinfo{journal}{Phys. Rev.} \textbf{\bibinfo{volume}{D77}},
  \bibinfo{pages}{015019} (\bibinfo{year}{2008}), \eprint{0709.4519}.

\bibitem[{\citenamefont{Dabelstein}(1995)}]{Dabelstein:1995js}
\bibinfo{author}{\bibfnamefont{A.}~\bibnamefont{Dabelstein}},
  \bibinfo{journal}{Nucl. Phys.} \textbf{\bibinfo{volume}{B456}},
  \bibinfo{pages}{25} (\bibinfo{year}{1995}), \eprint{hep-ph/9503443}.

\bibitem[{\citenamefont{Hall et~al.}(1994)\citenamefont{Hall, Rattazzi, and
  Sarid}}]{Hall:1993gn}
\bibinfo{author}{\bibfnamefont{L.~J.} \bibnamefont{Hall}},
  \bibinfo{author}{\bibfnamefont{R.}~\bibnamefont{Rattazzi}}, \bibnamefont{and}
  \bibinfo{author}{\bibfnamefont{U.}~\bibnamefont{Sarid}},
  \bibinfo{journal}{Phys. Rev.} \textbf{\bibinfo{volume}{D50}},
  \bibinfo{pages}{7048} (\bibinfo{year}{1994}), \eprint{hep-ph/9306309}.

\bibitem[{\citenamefont{Carena et~al.}(2000)\citenamefont{Carena, Garcia,
  Nierste, and Wagner}}]{Carena:1999py}
\bibinfo{author}{\bibfnamefont{M.~S.} \bibnamefont{Carena}},
  \bibinfo{author}{\bibfnamefont{D.}~\bibnamefont{Garcia}},
  \bibinfo{author}{\bibfnamefont{U.}~\bibnamefont{Nierste}}, \bibnamefont{and}
  \bibinfo{author}{\bibfnamefont{C.~E.~M.} \bibnamefont{Wagner}},
  \bibinfo{journal}{Nucl. Phys.} \textbf{\bibinfo{volume}{B577}},
  \bibinfo{pages}{88} (\bibinfo{year}{2000}), \eprint{hep-ph/9912516}.

\bibitem[{\citenamefont{Guasch et~al.}(2003)\citenamefont{Guasch, Hafliger, and
  Spira}}]{Guasch:2003cv}
\bibinfo{author}{\bibfnamefont{J.}~\bibnamefont{Guasch}},
  \bibinfo{author}{\bibfnamefont{P.}~\bibnamefont{Hafliger}}, \bibnamefont{and}
  \bibinfo{author}{\bibfnamefont{M.}~\bibnamefont{Spira}},
  \bibinfo{journal}{Phys. Rev.} \textbf{\bibinfo{volume}{D68}},
  \bibinfo{pages}{115001} (\bibinfo{year}{2003}), \eprint{hep-ph/0305101}.

\bibitem[{\citenamefont{Haber et~al.}(2001)}]{Haber:2000kq}
\bibinfo{author}{\bibfnamefont{H.~E.} \bibnamefont{Haber}}
  \bibnamefont{et~al.}, \bibinfo{journal}{Phys. Rev.}
  \textbf{\bibinfo{volume}{D63}}, \bibinfo{pages}{055004}
  (\bibinfo{year}{2001}), \eprint{hep-ph/0007006}.

\bibitem[{\citenamefont{Harlander and Kilgore}(2003)}]{Harlander:2003ai}
\bibinfo{author}{\bibfnamefont{R.~V.} \bibnamefont{Harlander}}
  \bibnamefont{and} \bibinfo{author}{\bibfnamefont{W.~B.}
  \bibnamefont{Kilgore}}, \bibinfo{journal}{Phys. Rev.}
  \textbf{\bibinfo{volume}{D68}}, \bibinfo{pages}{013001}
  (\bibinfo{year}{2003}), \eprint{hep-ph/0304035}.

\bibitem[{\citenamefont{Heinemeyer et~al.}(2005)\citenamefont{Heinemeyer,
  Hollik, Rzehak, and Weiglein}}]{Heinemeyer:2004xw}
\bibinfo{author}{\bibfnamefont{S.}~\bibnamefont{Heinemeyer}},
  \bibinfo{author}{\bibfnamefont{W.}~\bibnamefont{Hollik}},
  \bibinfo{author}{\bibfnamefont{H.}~\bibnamefont{Rzehak}}, \bibnamefont{and}
  \bibinfo{author}{\bibfnamefont{G.}~\bibnamefont{Weiglein}},
  \bibinfo{journal}{Eur. Phys. J.} \textbf{\bibinfo{volume}{C39}},
  \bibinfo{pages}{465} (\bibinfo{year}{2005}), \eprint{hep-ph/0411114}.

\bibitem[{\citenamefont{Brignole et~al.}(2002)\citenamefont{Brignole, Degrassi,
  Slavich, and Zwirner}}]{Brignole:2002bz}
\bibinfo{author}{\bibfnamefont{A.}~\bibnamefont{Brignole}},
  \bibinfo{author}{\bibfnamefont{G.}~\bibnamefont{Degrassi}},
  \bibinfo{author}{\bibfnamefont{P.}~\bibnamefont{Slavich}}, \bibnamefont{and}
  \bibinfo{author}{\bibfnamefont{F.}~\bibnamefont{Zwirner}},
  \bibinfo{journal}{Nucl. Phys.} \textbf{\bibinfo{volume}{B643}},
  \bibinfo{pages}{79} (\bibinfo{year}{2002}), \eprint{hep-ph/0206101}.

\bibitem[{\citenamefont{Noth and Spira}(2010)}]{Noth:2010jy}
\bibinfo{author}{\bibfnamefont{D.}~\bibnamefont{Noth}} \bibnamefont{and}
  \bibinfo{author}{\bibfnamefont{M.}~\bibnamefont{Spira}}
  (\bibinfo{year}{2010}), \eprint{1001.1935}.

\bibitem[{\citenamefont{Noth and Spira}(2008)}]{Noth:2008tw}
\bibinfo{author}{\bibfnamefont{D.}~\bibnamefont{Noth}} \bibnamefont{and}
  \bibinfo{author}{\bibfnamefont{M.}~\bibnamefont{Spira}},
  \bibinfo{journal}{Phys. Rev. Lett.} \textbf{\bibinfo{volume}{101}},
  \bibinfo{pages}{181801} (\bibinfo{year}{2008}), \eprint{0808.0087}.

\bibitem[{\citenamefont{Heinemeyer et~al.}(2000)\citenamefont{Heinemeyer,
  Hollik, and Weiglein}}]{Heinemeyer:1998yj}
\bibinfo{author}{\bibfnamefont{S.}~\bibnamefont{Heinemeyer}},
  \bibinfo{author}{\bibfnamefont{W.}~\bibnamefont{Hollik}}, \bibnamefont{and}
  \bibinfo{author}{\bibfnamefont{G.}~\bibnamefont{Weiglein}},
  \bibinfo{journal}{Comput. Phys. Commun.} \textbf{\bibinfo{volume}{124}},
  \bibinfo{pages}{76} (\bibinfo{year}{2000}), \eprint{hep-ph/9812320}.

\bibitem[{\citenamefont{Degrassi et~al.}(2003)\citenamefont{Degrassi,
  Heinemeyer, Hollik, Slavich, and Weiglein}}]{Degrassi:2002fi}
\bibinfo{author}{\bibfnamefont{G.}~\bibnamefont{Degrassi}},
  \bibinfo{author}{\bibfnamefont{S.}~\bibnamefont{Heinemeyer}},
  \bibinfo{author}{\bibfnamefont{W.}~\bibnamefont{Hollik}},
  \bibinfo{author}{\bibfnamefont{P.}~\bibnamefont{Slavich}}, \bibnamefont{and}
  \bibinfo{author}{\bibfnamefont{G.}~\bibnamefont{Weiglein}},
  \bibinfo{journal}{Eur. Phys. J.} \textbf{\bibinfo{volume}{C28}},
  \bibinfo{pages}{133} (\bibinfo{year}{2003}), \eprint{hep-ph/0212020}.

\bibitem[{\citenamefont{Heinemeyer et~al.}(1999)\citenamefont{Heinemeyer,
  Hollik, and Weiglein}}]{Heinemeyer:1998np}
\bibinfo{author}{\bibfnamefont{S.}~\bibnamefont{Heinemeyer}},
  \bibinfo{author}{\bibfnamefont{W.}~\bibnamefont{Hollik}}, \bibnamefont{and}
  \bibinfo{author}{\bibfnamefont{G.}~\bibnamefont{Weiglein}},
  \bibinfo{journal}{Eur. Phys. J.} \textbf{\bibinfo{volume}{C9}},
  \bibinfo{pages}{343} (\bibinfo{year}{1999}), \eprint{hep-ph/9812472}.

\bibitem[{\citenamefont{Dittmaier et~al.}(2007)\citenamefont{Dittmaier, Kramer,
  Muck, and Schluter}}]{Dittmaier:2006cz}
\bibinfo{author}{\bibfnamefont{S.}~\bibnamefont{Dittmaier}},
  \bibinfo{author}{\bibfnamefont{M.}~\bibnamefont{Kramer}},
  \bibinfo{author}{\bibfnamefont{A.}~\bibnamefont{Muck}}, \bibnamefont{and}
  \bibinfo{author}{\bibfnamefont{T.}~\bibnamefont{Schluter}},
  \bibinfo{journal}{JHEP} \textbf{\bibinfo{volume}{03}}, \bibinfo{pages}{114}
  (\bibinfo{year}{2007}), \eprint{hep-ph/0611353}.

\bibitem[{\citenamefont{Carena et~al.}(1994)\citenamefont{Carena, Olechowski,
  Pokorski, and Wagner}}]{Carena:1994bv}
\bibinfo{author}{\bibfnamefont{M.~S.} \bibnamefont{Carena}},
  \bibinfo{author}{\bibfnamefont{M.}~\bibnamefont{Olechowski}},
  \bibinfo{author}{\bibfnamefont{S.}~\bibnamefont{Pokorski}}, \bibnamefont{and}
  \bibinfo{author}{\bibfnamefont{C.~E.~M.} \bibnamefont{Wagner}},
  \bibinfo{journal}{Nucl. Phys.} \textbf{\bibinfo{volume}{B426}},
  \bibinfo{pages}{269} (\bibinfo{year}{1994}), \eprint{hep-ph/9402253}.

\bibitem[{\citenamefont{Dittmaier et~al.}(2009)\citenamefont{Dittmaier, Kramer,
  Spira, and Walser}}]{Dittmaier:2009np}
\bibinfo{author}{\bibfnamefont{S.}~\bibnamefont{Dittmaier}},
  \bibinfo{author}{\bibfnamefont{M.}~\bibnamefont{Kramer}},
  \bibinfo{author}{\bibfnamefont{M.}~\bibnamefont{Spira}}, \bibnamefont{and}
  \bibinfo{author}{\bibfnamefont{M.}~\bibnamefont{Walser}}
  (\bibinfo{year}{2009}), \eprint{0906.2648}.

\bibitem[{\citenamefont{Berger et~al.}(2005)\citenamefont{Berger, Han, Jiang,
  and Plehn}}]{Berger:2003sm}
\bibinfo{author}{\bibfnamefont{E.~L.} \bibnamefont{Berger}},
  \bibinfo{author}{\bibfnamefont{T.}~\bibnamefont{Han}},
  \bibinfo{author}{\bibfnamefont{J.}~\bibnamefont{Jiang}}, \bibnamefont{and}
  \bibinfo{author}{\bibfnamefont{T.}~\bibnamefont{Plehn}},
  \bibinfo{journal}{Phys. Rev.} \textbf{\bibinfo{volume}{D71}},
  \bibinfo{pages}{115012} (\bibinfo{year}{2005}), \eprint{hep-ph/0312286}.

\bibitem[{\citenamefont{Hofer et~al.}(2009)\citenamefont{Hofer, Nierste, and
  Scherer}}]{Hofer:2009xb}
\bibinfo{author}{\bibfnamefont{L.}~\bibnamefont{Hofer}},
  \bibinfo{author}{\bibfnamefont{U.}~\bibnamefont{Nierste}}, \bibnamefont{and}
  \bibinfo{author}{\bibfnamefont{D.}~\bibnamefont{Scherer}},
  \bibinfo{journal}{JHEP} \textbf{\bibinfo{volume}{10}}, \bibinfo{pages}{081}
  (\bibinfo{year}{2009}), \eprint{0907.5408}.

\bibitem[{\citenamefont{Accomando et~al.}(2011)\citenamefont{Accomando,
  Chachamis, Fugel, Spira, and Walser}}]{Accomando:2011jy}
\bibinfo{author}{\bibfnamefont{E.}~\bibnamefont{Accomando}},
  \bibinfo{author}{\bibfnamefont{G.}~\bibnamefont{Chachamis}},
  \bibinfo{author}{\bibfnamefont{F.}~\bibnamefont{Fugel}},
  \bibinfo{author}{\bibfnamefont{M.}~\bibnamefont{Spira}}, \bibnamefont{and}
  \bibinfo{author}{\bibfnamefont{M.}~\bibnamefont{Walser}}
  (\bibinfo{year}{2011}), \eprint{1103.4283}.

\bibitem[{\citenamefont{Berge et~al.}(2007)\citenamefont{Berge, Hollik, Mosle,
  and Wackeroth}}]{Berge:2007dz}
\bibinfo{author}{\bibfnamefont{S.}~\bibnamefont{Berge}},
  \bibinfo{author}{\bibfnamefont{W.}~\bibnamefont{Hollik}},
  \bibinfo{author}{\bibfnamefont{W.~M.} \bibnamefont{Mosle}}, \bibnamefont{and}
  \bibinfo{author}{\bibfnamefont{D.}~\bibnamefont{Wackeroth}},
  \bibinfo{journal}{Phys. Rev.} \textbf{\bibinfo{volume}{D76}},
  \bibinfo{pages}{034016} (\bibinfo{year}{2007}), \eprint{hep-ph/0703016}.

\bibitem[{\citenamefont{Nason et~al.}(1988)\citenamefont{Nason, Dawson, and
  Ellis}}]{Nason:1987xz}
\bibinfo{author}{\bibfnamefont{P.}~\bibnamefont{Nason}},
  \bibinfo{author}{\bibfnamefont{S.}~\bibnamefont{Dawson}}, \bibnamefont{and}
  \bibinfo{author}{\bibfnamefont{R.~K.} \bibnamefont{Ellis}},
  \bibinfo{journal}{Nucl. Phys.} \textbf{\bibinfo{volume}{B303}},
  \bibinfo{pages}{607} (\bibinfo{year}{1988}).

\bibitem[{\citenamefont{Gunion et~al.}(1988)\citenamefont{Gunion, Haber, and
  Sher}}]{Gunion:1987qv}
\bibinfo{author}{\bibfnamefont{J.~F.} \bibnamefont{Gunion}},
  \bibinfo{author}{\bibfnamefont{H.~E.} \bibnamefont{Haber}}, \bibnamefont{and}
  \bibinfo{author}{\bibfnamefont{M.}~\bibnamefont{Sher}},
  \bibinfo{journal}{Nucl. Phys.} \textbf{\bibinfo{volume}{B306}},
  \bibinfo{pages}{1} (\bibinfo{year}{1988}).

\bibitem[{\citenamefont{Nadolsky et~al.}(2008)}]{Nadolsky:2008zw}
\bibinfo{author}{\bibfnamefont{P.~M.} \bibnamefont{Nadolsky}}
  \bibnamefont{et~al.}, \bibinfo{journal}{Phys. Rev.}
  \textbf{\bibinfo{volume}{D78}}, \bibinfo{pages}{013004}
  (\bibinfo{year}{2008}), \eprint{0802.0007}.

\bibitem[{\citenamefont{Martin et~al.}(2009)\citenamefont{Martin, Stirling,
  Thorne, and Watt}}]{Martin:2009iq}
\bibinfo{author}{\bibfnamefont{A.~D.} \bibnamefont{Martin}},
  \bibinfo{author}{\bibfnamefont{W.~J.} \bibnamefont{Stirling}},
  \bibinfo{author}{\bibfnamefont{R.~S.} \bibnamefont{Thorne}},
  \bibnamefont{and} \bibinfo{author}{\bibfnamefont{G.}~\bibnamefont{Watt}},
  \bibinfo{journal}{Eur. Phys. J.} \textbf{\bibinfo{volume}{C63}},
  \bibinfo{pages}{189} (\bibinfo{year}{2009}), \eprint{0901.0002}.

\end{thebibliography}

\end{document}